\documentclass[twocolumn]{aastex631}

\usepackage{graphicx} \usepackage{color}
\usepackage{float}
\usepackage{tabularx}
\usepackage{comment}
\usepackage{natbib}
\usepackage{amsmath}
\usepackage{booktabs}
\usepackage{array}
\usepackage{hyperref}
\graphicspath{{./}{figures/}}

\newcommand{\rion}[2]{{\ensuremath{\mbox{\rm #1$\,${\sc\expandafter{\romannumeral#2\relax}}}}}}
\newcommand{\HII}{\rion{H}{2}}
\newcommand{\NII}{\rion{N}{2}}

\begin{document}

\title{\textit{JWST} Reveals a Candidate Supermassive Black Hole Binary at $z=4.3$ in the Brightest Sub-millimeter Galaxy in COSMOS-Web}

\author[0000-0002-6149-8178]{Jed McKinney}
\email{jed.mckinney.astro@gmail.com}
\altaffiliation{NASA Hubble Fellow}
\affiliation{Department of Astronomy, The University of Texas at Austin, Austin, TX 78712, USA}
\affiliation{Cosmic Frontier Center, The University of Texas at Austin, Austin, TX 78712, USA}

\author[0000-0003-4242-8606]{Ansh Gupta}
\altaffiliation{NSF Graduate Research Fellow}
\affiliation{Department of Astronomy, The University of Texas at Austin, Austin, TX 78712, USA}
\affiliation{Cosmic Frontier Center, The University of Texas at Austin, Austin, TX 78712, USA}

\author[0000-0002-8984-0465]{Julian B. Mu\~noz}
\affiliation{Department of Astronomy, The University of Texas at Austin, Austin, TX 78712, USA}
\affiliation{Cosmic Frontier Center, The University of Texas at Austin, Austin, TX 78712, USA}

\author[0000-0002-0302-2577]{John Chisholm}
\affiliation{Department of Astronomy, The University of Texas at Austin, Austin, TX 78712, USA}
\affiliation{Cosmic Frontier Center, The University of Texas at Austin, Austin, TX 78712, USA}

\author[0000-0002-0930-6466]{Caitlin M. Casey}
\affiliation{Department of Physics, University of California Santa Barbara, Santa Barbara, CA, USA}

\author[0000-0001-8169-7249]{Stephanie M. Urbano Stawinski}
\affiliation{Department of Physics, University of California Santa Barbara, Santa Barbara, CA, USA}

\author[0000-0003-3881-1397]{Olivia Cooper}
\altaffiliation{NSF Astronomy \& Astrophysics Postdoctoral Fellow}
\affiliation{Department for Astrophysical and Planetary Science, University of Colorado, Boulder, CO 80309, USA}

\author[0000-0003-3216-7190]{Erini Lambrides}\altaffiliation{NPP Fellow}
\affiliation{NASA-Goddard Space Flight Center, Code 662, Greenbelt, MD, 20771, USA}

\author[0000-0003-3596-8794]{Hollis Akins}
\altaffiliation{NSF Graduate Research Fellow}
\affiliation{Department of Astronomy, The University of Texas at Austin, Austin, TX 78712, USA}
\affiliation{Cosmic Frontier Center, The University of Texas at Austin, Austin, TX 78712, USA}

\author[0000-0002-3560-8599]{Maximilien Franco}
\affiliation{Universit\'e Paris-Saclay, Universit\'e Paris Cit\'e, CEA, CNRS, AIM, 91191, Gif-sur-Yvette, France}

\author[0000-0001-7578-2412]{Archana Aravindan}
\affiliation{Department of Astronomy, The University of Texas at Austin, Austin, TX 78712, USA}
\affiliation{Cosmic Frontier Center, The University of Texas at Austin, Austin, TX 78712, USA}

\author[0000-0001-7201-5066]{Seiji Fujimoto}
\affiliation{David A. Dunlap Department of Astronomy and Astrophysics, University of Toronto, Toronto, ON M5S 3H4, Canada}
\affiliation{Dunlap Institute for Astronomy and Astrophysics, University of Toronto, Toronto, ON M5S 3H4, Canada}

\author[0000-0001-9840-4959]{Kohei Inayoshi}
\affiliation{Kavli Institute for Astronomy and Astrophysics, Peking University, Beijing 100871, China}

\author[0000-0002-9382-9832]{Andreas L. Faisst}
\affiliation{Caltech/IPAC, MS 314-6, 1200 E. California Blvd. Pasadena, CA 91125, USA}

\author[0000-0001-9187-3605]{Jeyhan S. Kartaltepe}
\affiliation{Laboratory for Multiwavelength Astrophysics, School of Physics and Astronomy, Rochester Institute of Technology, 84 Lomb Memorial Drive, Rochester, NY 14623, USA}

\author[0000-0002-9604-343X]{Michael Boylan-Kolchin}
\affiliation{Department of Astronomy, The University of Texas at Austin, Austin, TX 78712, USA}
\affiliation{Cosmic Frontier Center, The University of Texas at Austin, Austin, TX 78712, USA}
\affiliation{Texas Center for Cosmology and Astroparticle Physics, Weinberg Institute, The University of Texas at Austin, Austin, TX 78712, USA}

\begin{abstract}
We present \textit{JWST}/NIRSpec PRISM and G395M grating spectroscopy for AzTEC-1, a massive sub-mm bright galaxy at $z=4.34$ in the COSMOS extragalactic field. The PRISM spectrum reveals strong H$\alpha$, a significant Balmer break, and no H$\beta$ detection, indicating a $100-400$ Myr-old stellar population and high dust attenuation. BPT line ratios indicate the presence of an Active Galactic Nucleus (AGN). Decomposing narrow and broad line components, we recover broad, blueshifted H$\alpha$ with a velocity offset of $1245{\,\rm km\,s^{-1}}$ from the systemic narrow line velocity and with FWHM$\,\sim2500\,{\rm km\,s^{-1}}$. AzTEC-1's smooth morphology and stellar age is suggestive of a past merger-induced starburst period that would have brought in a second supermassive black hole, raising the possibility for a binary supermassive black hole system. In this scenario, we assume that the lower mass black hole hosts a broad line region orbiting a quiescent primary. Evidence for an extended outflow is not found in the 2D spectrum, NIRCam imaging, resolved ALMA observations of dust continuum, or CO, [\ion{C}{2}]$_{157\,\mu \rm m}$ and [\ion{N}{2}]$_{\rm 205\,\mu m}$ kinematics. AzTEC-1's high central gas mass surface density and dynamically unstable gas disk indicates that massive gas clouds external to the candidate binary SMBH's orbit might have played a role in stalling infall from $\mathcal{O}(10$ Myr) to $\mathcal{O}(100$ Myr) through dynamical torques, which has been theorized to occur in the nuclei of massive galaxies like AzTEC-1. If the supermassive black hole binary is confirmed, AzTEC-1 would be an excellent laboratory into the astrophysics driving low-frequency gravitational wave detections. 
\end{abstract}


\section{Introduction \label{sec:intro}}

Galaxy mergers help assemble the most massive, active galaxies at early cosmic times and fuel central supermassive black hole (SMBH) growth \citep[e.g.,][]{Hopkins2006}. Sometime after the galaxy merger occurs, the pair of SMBHs might lose sufficient angular momentum to form a binary at which point stochastic interactions with the stellar and gaseous environment might remove enough energy for them to merge into a single SMBH \citep{Frank1976,Begelman1980,Armitage2002,Milosavljevic2005,Cuadra2009,Hayasaki2009,Khan2016}. Spin and mass asymmetries can lead to anisotropic gravitational waves around coalescence that can impart a velocity kick to the remnant \citep{Bekenstein1973}, which in some cases is sufficient for the merged SMBH to escape its host galaxy \citep{Redmount1989,Campanelli2007,Gonzalez2007}. Pulsar Timing Arrays have detected a low-frequency gravitational wave background thought to originate from SMBH mergers at $z<1$ \citep{Agazie2023,Reardon2023,Xu2023,EPTA2024}. Facilities like the Laser Interferometer Space Antenna (LISA) will be able to detect the gravitational wave signatures from individual merging binaries at lower total masses and up to higher redshifts \citep{Arun2007,Katz2019,LISA2025}, but direct electromagnetic constraint on this process has yet to be observed at $z\gtrsim1.5$ \citep{Liao2021}. 

The abundance of binary SMBHs and their distribution across mass and redshift is a function of galaxy stellar mass functions, galaxy merger rates, dynamical evolution with respect to stellar and gas-rich environments, and gravitational wave energy loss \citep[e.g.,][]{Lu2025}. Binary SMBHs encode information on the kpc-to-pc scale orbital evolution needed to produce the binary, as well as the assembly of SMBHs more generally and their scaling with host galaxy properties. SMBHs spend most of their active lifetime heavily obscured by gas and dust \citep{Hopkins2006}, which can make observing the electromagnetic counterparts to binary SMBHs and SMBH mergers difficult. Naturally most candidate binary SMBH systems to-date have been discovered at relatively low-redshift $(z<1)$ in optically-luminous galaxies \citep{Komossa2008,Bogdanovic2009,Boroson2009,Dotti2009,Shields2009,Robinson2010,Decarli2010,Barrows2011,Steinhardt2012,Tsai2013,Civano2012,Barrows2025}, with higher redshift candidates having projected physical separations $\gtrsim1$ kpc \citep[e.g.,][]{Orosz2013,Chen2022vodka,Li2025,Pfeifle2025,Dadiani2026}. Dust-obscured galaxies bright at mm/sub-mm wavelengths are thought to grow through major mergers \citep{Hopkins2008}, and they host some of the most massive and actively growing SMBHs \citep{Wang2013,Umehata2015,Rujopakarn2016,Ueda2018,Stach2019,Uematsu2023}, making them a natural galaxy population to chart black hole evolution through. However, the high-resolution rest-frame UV/optical spectroscopic signatures needed to test for binary SMBHs were historically difficult to access through high levels of dust attenuation \citep{Casey2014b,Lambrides2020}. The sensitivity of \textit{JWST} has now opened up traditional diagnostics like [\ion{N}{2}], H$\alpha$, and [\ion{O}{3}] in high-redshift dusty star-forming galaxies \citep[e.g.,][]{Cooper2025}.

AzTEC-1 (RA, Dec$=$149.928566, 2.493960 [deg]) was discovered in the COSMOS field \citep{Scoville2007} as the brightest source detected by the first AzTEC 1.1mm survey on the James Clerk Maxwell Telescope \citep{Scott2008}.\footnote{Brighter sources were later discovered in subsequent AzTEC/JCMT surveys of COSMOS that followed similar naming conventions \citep[e.g.,][]{Aretxaga2011}. In particular, the source in this paper should not be confused with AzTEC/C1, the brightest 1.1mm source in COSMOS to-date.} A spectroscopic redshift from CO(5-4) and CO(4-3) was confirmed by \cite{Yun2015} at $z = 4.3420\pm0.0004$ using the Large Millimeter Telescope. \cite{Iono2016} performed high-resolution $\theta\sim0.03^{\prime\prime}$ ALMA $860\,\mu$m imaging and resolved the inner kpc into multiple clumps, including two within the innermost 100 pc. \cite{Tadaki2018} use CO(4-3) velocity widths to resolve the Toomre $Q$ parameter, finding that the system contains a dynamically unstable disk. \cite{Tadaki2019} observe several more emission lines with ALMA including [\ion{C}{2}] $157\,\mu$m and [\ion{O}{3}] $88\,\mu$m, and report that the neutral, ionized, and gas associated with photo-dissociation regions (PDRs) across the extent of the galaxy all have similar kinematic properties, and that the system is rotationally supported. \cite{Tadaki2020} uncover residual non-corotating gas possibly driven by a past merger(s). No obvious evidence for an obscured Active Galactic Nucleus (AGN) is found, but nominally AzTEC-1's high IR luminosity ($9\times\sim10^{12}\,L_\odot$,  \citealt{Smolcic2011}) allows for some degree of AGN-heating of cold dust emission as has been suggested by simulations \citep{McKinney2021agn,Bardati2026}. 

In this paper we present \textit{JWST}/NIRSpec PRISM and G395M spectroscopy in AzTEC-1 which detect a rich array of features including a prominent Balmer break, UV continuum, Ly$\alpha$, and nebular emission lines. The observed [\ion{N}{2}]$/{\rm H\alpha}$ ratio indicates the presence of an AGN, and most notably, the shape of the H$\alpha$ profile is highly asymmetric due to a kinematically distinct broad line component offset from the narrow line emission. Similar profiles have been observed in galaxies at lower redshifts and are thought to arise from supermassive black hole (SMBH) binaries, kicked SMBHs, or outflows \citep[e.g.,][]{Komossa2008,Bogdanovic2009,Harrison2012,Civano2012,Eracleous2012,Ju2013,Ubler2023}. 
Section \ref{sec:data} details our data reduction and analysis, the results from which we present in Section \ref{sec:results}. We interpret these observations in Section \ref{sec:disc} as a candidate supermassive binary black hole,  a recoiling SMBH that has received a kick relatively recently, or an ionized outflow which is disfavored. Section \ref{sec:conc} summarizes our conclusions. 

Throughout this work we assume a $\Lambda$CDM cosmology with $H_0=70$\,km\,s$^{-1}$\,Mpc$^{-1}$, $\Omega_m=0.3$, $\Omega_\Lambda=0.7$, a Chabrier initial mass function (IMF, \citealt{Chabrier2003}), and the AB magnitude system \citep{Oke1974}. 

\begin{figure}
    \centering
    \includegraphics[width=\linewidth]{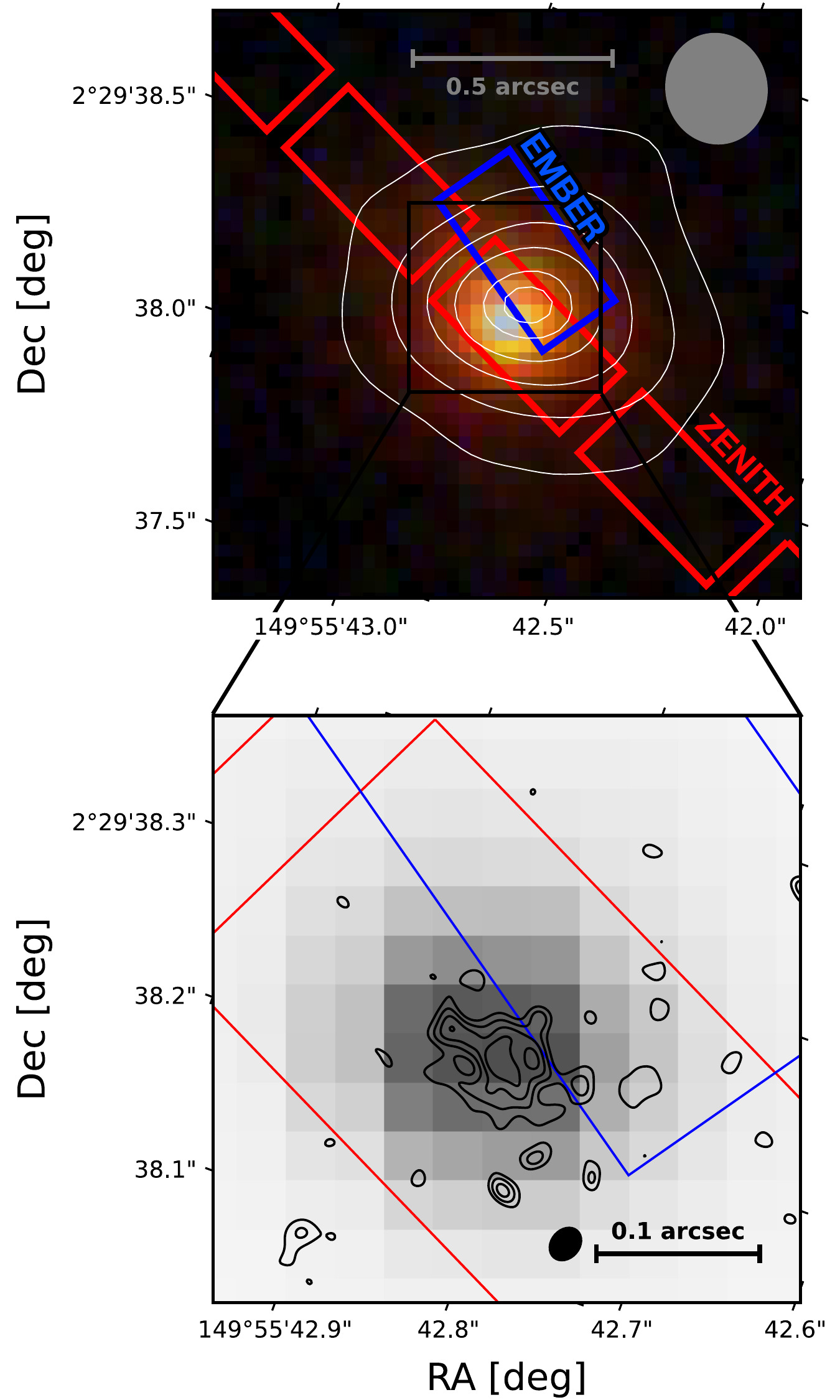}
    \caption{\textit{JWST}/NIRSpec slit apertures compared with NIRCam morphology and ALMA dust continuum. The red and blue regions on both panels indicate the ZENITH (PRISM) and EMBER (PRISM+G395M) slit placements respectively. Scale bars are included for reference, and ellipses represent the primary beam of the ALMA data over-plotted as contours. (\textit{Top}) NIRCam RGB map constructed from F444W/F277W/F150W in the background with $\theta\sim0.3^{\prime\prime}$ resolution ALMA observations (contours are $3\sigma,\,10\sigma,\,20\sigma,\,30\sigma,\,40\sigma,\,45\sigma$) which has a centroid within both program's slits. (\textit{Bottom}) F356W imaging in the background, which overlaps [\ion{N}{2}]$\,+\,$H$\alpha$ at $z=4.3$, compared to $\theta\sim0.015^{\prime\prime}$ resolution ALMA observations (contours are $3\sigma-6\sigma$). The high-resolution ALMA data \citep{Iono2016} resolves compact clumps which overlap the F356W centroid.}
    \label{fig:rgb}
\end{figure}

\begin{figure}[h!]
    \centering
    \includegraphics[width=\linewidth]{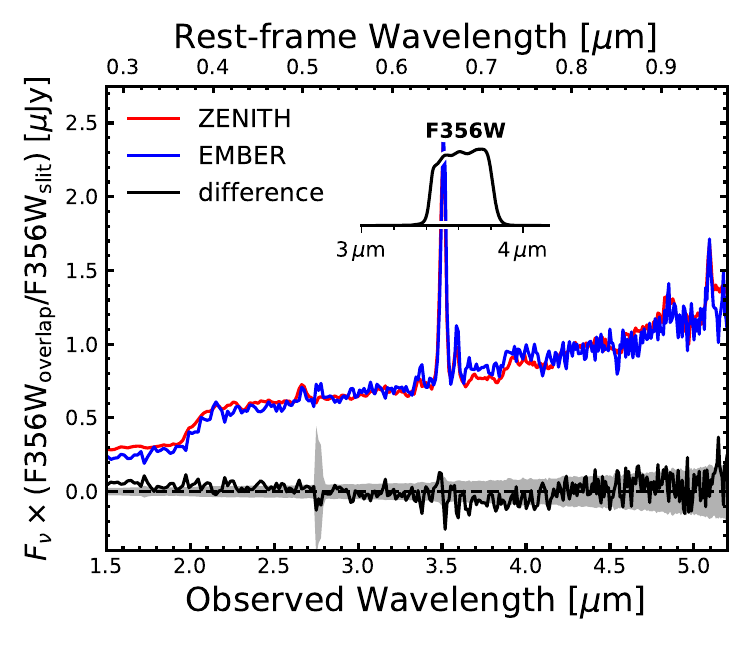}
    \caption{PRISM spectra multiplied by the fraction of F356W emission in the ZENITH/EMBER overlap region relative to the total F356W flux measured in the slit aperture from each program. The difference between the scaled ZENITH and EMBER spectra (black line) is consistent with zero, indicating that the difference in total flux is an aperture effect arising from the offset slit placements as shown in Figure \ref{fig:rgb}. The inset panel shows the filter transmission curve for F356W for reference.}
    \label{fig:prism_compare}
\end{figure}

\begin{figure*}
    \centering
    \includegraphics[width=\linewidth]{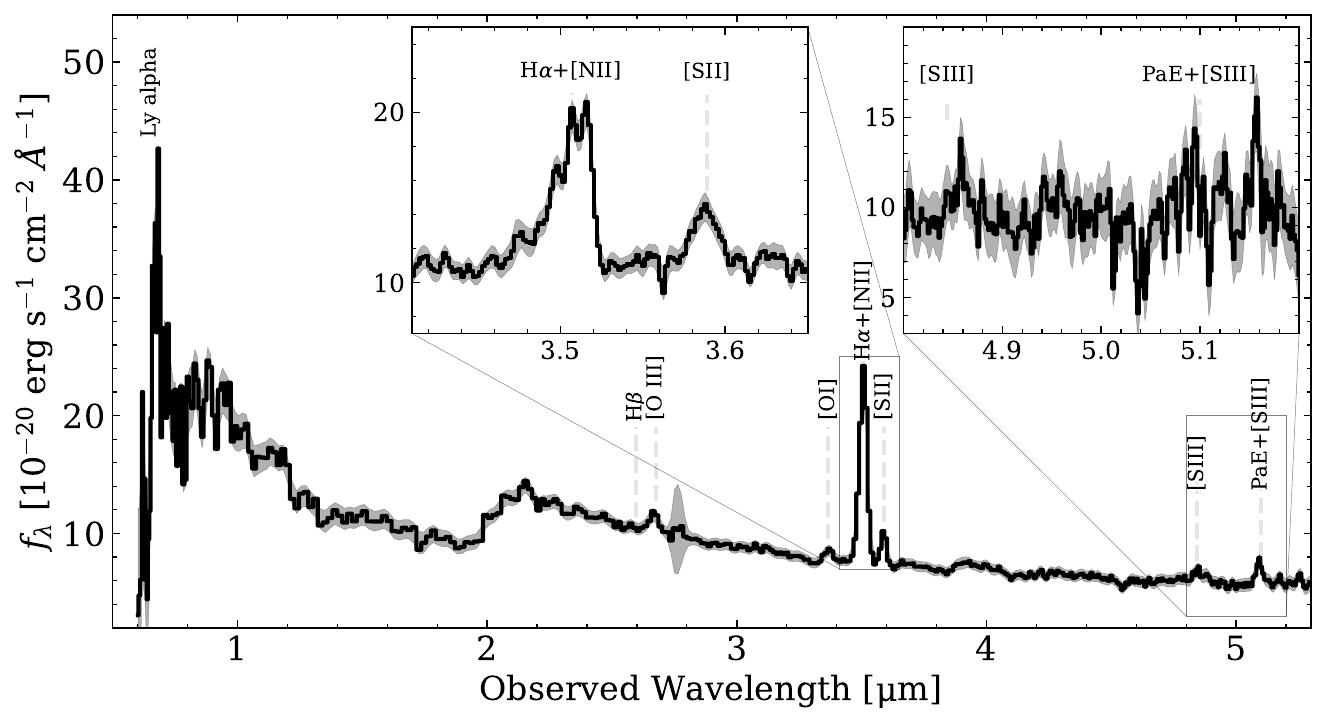}
    \caption{AzTEC-1's combined and slit-loss corrected NIRSPec/PRISM spectrum from EMBER (29 min) and ZENITH (1.1hr) for a total of $1.6$ hours of exposure time. The prism spectra from both programs have been individually slit-loss corrected by normalizing to \textit{HST} and \textit{JWST} photometry, and then combined.  Inset panels show the 18 min G395M spectrum from EMBER capturing the H$\alpha+$[\ion{N}{2}] complex, [\ion{S}{2}] doublet, and [\ion{S}{3}]. }
    \label{fig:spec}
\end{figure*}

\section{\textit{JWST} Data Reduction and Analysis\label{sec:data}}
We obtain \textit{JWST} NIRSpec micro-shutter assembly (MSA) spectra for AzTEC-1 from EMBER (Cycle 4 \#7076, PI Akins) and ZENITH (Cycle 4 \#7417, PI Casey). EMBER obtained a 29 minute PRISM exposure of the target on 04/09/2026 beginning 19:14 UTC, followed by an 18 minute exposure of the target in G395M beginning 20:52 UTC. ZENITH obtained a $1.1$ hr PRISM exposure on 04/20/2026 at 00:52 UTC. Figure \ref{fig:rgb} shows the slitlet placements for both programs relative to NIRCam and ALMA imaging. We retrieved the spectra from CAMPFIRE (the COSMOS Archive of MultiPle-Field Internal Reductions \& Extractions, Akins et al., in prep.)\footnote{\url{https://campfire.hollisakins.com/}} which processes NIRSpec spectra with the standard \textit{JWST} calibration pipeline with custom processing steps to handle multi-component background models at the rate file level, characterize detector noise, and extract spectra using the JADES DR4 empirical wavelength zero-point correction \citep{Scholtz2025}. 

As can be seen on Figure \ref{fig:rgb} the EMBER and ZENITH observations have similar slit orientation angles but different center placements. The ZENITH slits, through which only a PRISM spectrum was measured, use 5 NIRSpec slitlets and overlap the NIRCam photometric centroid. The EMBER slit, through which both a PRISM and G395M spectrum were taken, is comprised of a single slitlet that does not encase the photometric center. These different slit placements arise from AzTEC-1's nature as a filler target in both programs. We note that the astrometry of these slits are accurate to $\approx0.05^{\prime\prime}$. To test if there are differences between the two programs' PRISM spectra that could arise from different spatial sampling, we measure aperture photometry through the slit configurations for both programs using the F356W image, as well as in the overlap region between the two. We then calculate the fraction of F356W flux in the overlap region relative to the total slit aperture flux, and scale each PRISM spectrum by this normalization factor. As shown in Figure \ref{fig:prism_compare} the normalized EMBER and ZENITH spectra are consistent with one-another within uncertainties. We therefore proceed to combine the two spectra which are dominated by the same compact nuclear source emission.

To stack the two independent PRISM spectra, we first derive a wavelength-dependent normalization factor for each by convolving the spectrum with NIRCam photometry that spans the full wavelength range. We then combine the spectra using a flux conserving resampling algorithm implemented in \texttt{specutils} to obtain an effective exposure time of $t_{\rm exp}=95$ min. Finally, we normalize the G395M spectrum to the integrated flux of the unresolved H$\alpha\,+\,$[\ion{N}{2}] complex in the PRISM spectrum. The final combined PRISM and normalized G395M spectrum are shown in Figure \ref{fig:spec}. In the PRISM spectrum we detect a prominent Balmer break, Ly$\alpha$, the [\ion{O}{3}]$\lambda4932,\,\lambda4960$\AA\ doublet, [\ion{O}{1}]$\lambda6300$\AA, the [\ion{N}{2}]$\lambda6545,\,\lambda6585$\AA\ doublet blended with H$\alpha$, the [\ion{S}{2}]$\lambda6718,\,\lambda6732$\AA\ doublet, and the [\ion{S}{3}]$\lambda9068\,\lambda 9545$\AA\ doublet blended with Pa$\epsilon$. In the G395M spectrum we resolve [\ion{N}{2}] and H$\alpha$ as well as [\ion{S}{2}]$\lambda6718,\,\lambda6732$\AA, but [\ion{S}{3}] and Pa$\epsilon$ are marginally detected (SNR$<5$) while [\ion{O}{1}]$\lambda6365$\AA\ is undetected.  

\subsection{\textit{JWST} Imaging}

\textit{JWST} imaging data over our target comes from the cycle 1  program COSMOS-Web, a 0.54 deg$^2$ survey using NIRCam and MIRI (PID \#1727, PIs Kartaltepe \& Casey, \citealt{cosmos-web}) centered in the middle of the 2 deg$^2$ COSMOS survey field \cite{Scoville2007}. The COSMOS-Web NIRCam filters include F115W, F150W, F277W, and F444W that reach $5\sigma$ depths between $\sim27-28$ mag \citep{Franco2026}, and the MIRI F770W data reaches $5\sigma=25.51$ mag \citep{Harish2025}. AzTEC-1 does not overlap the MIRI coverage in COSMOS from COSMOS-Web. We additionally use data from COSMOS-3D (PID \#5893, PI Kakiichi) which provides imaging in F200W and F356W at depths of 28.3 mag and 28.6 mag respectively in the vicinity of AzTEC-1. The data reduction for these filters will be described in Akins et al., in prep., and follows the COSMOS-Web reduction described in \citealt{Franco2026} now applied to the new COSMOS-3D data. 

\subsection{ALMA Observations}
The results of deep and spatially resolved ALMA observations of AzTEC-1 have been discussed extensively by \cite{Iono2016}, and \cite{Tadaki2018,Tadaki2019,Tadaki2020}. Throughout this paper we refer to the results presented therein. For our own comparative imaging analysis we download Band 7 data from the ALMA Science Archive for the following programs: 2012.1.00978.S (PI Karim, $\theta=0.3^{\prime\prime}$), 2015.1.01345.S (PI Iono, $\theta=0.015^{\prime\prime}$), and 2017.1.00127.S (PI Iono, $\theta=0.09^{\prime\prime}$). We create two images from these observations, first a combined continuum map after downweighting the visibilities from 2012.1.00978.S and 2017.1.00127.S by factors of 10 and 6 respectively to account for differences in spatial resolution relative to 2012.1.00978.S. This map is dominated by the large scale structure shown in Figure \ref{fig:rgb} (\textit{Top}). Next we create a resolved continuum map generally following \cite{Iono2016} by applying a $0.01^{\prime\prime}$ $uv-$taper to the high resolution data from 2015.1.01345.S, which can be seen in Figure \ref{fig:rgb} (\textit{Bottom}). For both maps we use Briggs weighting, $R=0.5$, and interactive masking during \texttt{tclean}. 

\section{Results\label{sec:results}}

\begin{table}
	\centering
	\caption{AzTEC-1 Observed and Inferred Properties}
	\label{tab:source}
	\begin{tabular}{lr} 
		\hline
        Parameter  & Value/Units   \\
      \hline
         $z$ & $4.3419\pm0.0002$  \\  
         $\rm SFR_{IR}$\tablenotemark{$\dagger$} & $1169^{+11}_{-274}\,M_\odot\,\rm yr^{-1}$\\
         $t_{50}$\tablenotemark{a} & $314^{+78}_{-73}$ Myr \\
         $t_{90}$\tablenotemark{b} & $140^{+35}_{-35}$ Myr \\
         $f_{\rm AGN}/f_{\rm tot} (5500 {\rm \AA})$\tablenotemark{c} & $0.26^{+0.08}_{-0.07}$ \\
         $\mathrm{log}$ $Z_*/\mathrm{Z}_{\odot}$ & $0.12^{+0.08}_{-0.19}$ \\
         $\mathrm{log}$ $Z_{\rm NLR}/\mathrm{Z}_{\odot}$ & $0.38^{+0.01}_{-0.02}$ \\
         $\log\,M_*/M_\odot$ & $10.6\pm0.05$ \\ 
         $\log\,M_{\rm BH}/M_\odot$ & $8.2\pm0.5$ \\ 
      $A_V^{\rm stellar}$ & $1.6\pm0.2$ mag \\   
      $A_V^{\rm neb}$ & $4.2\pm1.2$ mag \\
        $L_{\rm H\alpha,\,narrow}$ & $2.03\pm0.18\times10^{10}\,L_\odot$ \\ 
        $L_{\rm H\alpha,\,broad}$ & $1.6\pm0.2\times10^{10}\,L_\odot$ \\ 
      FWHM(H$\alpha)_{\rm narrow}$ & $290^{+7}_{-16}\,{\rm km\,s^{-1}}$ \\
      FWHM(H$\alpha)_{\rm broad}$ & $2520^{+345}_{-367}\,{\rm km\,s^{-1}}$ \\ 
      FWHM([\ion{S}{2}])$_{\rm broad}$ & $1160^{+740}_{-370}\,{\rm km\,s^{-1}}$ \\ 
      $\Delta v_{\rm BLR,\,H\alpha}$ & $-1245^{+242}_{-210}\,{\rm km\,s^{-1}}$ \\
      $\Delta v_{\rm[N\,II]}$ & $-152^{+18}_{-17}\,{\rm km\,s^{-1}}$ \\ 
      $\Delta v_{\rm\,[S\,II]}$ & $-460^{+170}_{-180}\,{\rm km\,s^{-1}}$ \\
      $\log\,{\rm [N\,II]/H\alpha_{narrow}}$ & $0.23\pm0.03$\\
      $\log\,{\rm [O\,I]/H\alpha_{narrow}}$ & $-0.6\pm0.02$\\
      $\log\,{\rm [S\,II]/H\alpha_{narrow}}$ & $-0.16\pm0.04$\\
      $\log\,{\rm [O\,III]/[O\,I]}$ & $0.34\pm0.20$\\
      ${\rm [S\,II]6716\mathring{A}/[S\,II]6731\mathring{A}}$ & $1.1\pm0.4$\\
      $z_{\rm Ly\alpha}$ & $4.47\pm0.05$  \\  
      EW(Ly$\alpha)$ & $35\pm5$ \AA \\
      \hline
	\end{tabular}
    \tablecomments{\raggedright Reported luminosities have been corrected for dust attenuation. $\dagger$ From \cite{Tadaki2018}. $^a$ Look-back time at which $50\%$ of the stellar mass was formed. $^b$ Look-back time at which $90\%$ of the stellar mass was formed. $^c$ Fractional AGN contribution at rest-frame $5500$\AA. }
\end{table}

\begin{figure}
    \centering
    \includegraphics[width=\linewidth]{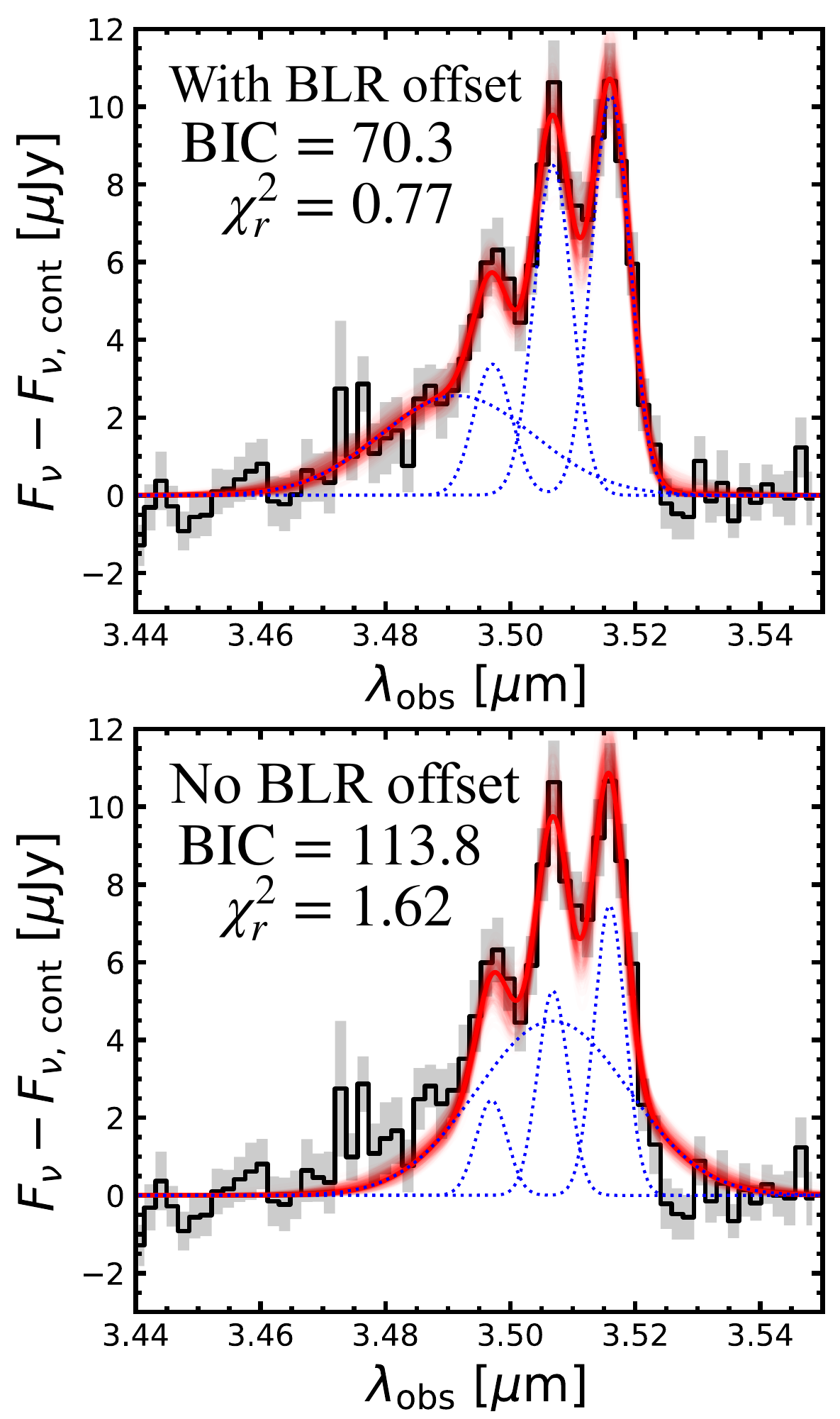}
    \caption{NIRSpec/G395M spectrum resolving the [N II], H$\alpha$ complex showing detections of both broad and narrow components. The G395M spectrum has been normalized to the total flux from the slit-loss corrected PRISM spectrum, and then continuum-subtracted. Red curves represent 100 random samples from fit posteriors, while the dashed blue lines show the individual line components from the most likely model. The \textit{Top} panel shows a multi-component fit where broad H$\alpha$ is fixed to the systemic velocity of the narrow H$\alpha$ line, whereas the \textit{Bottom} panel shows our fiducial model where the broad H$\alpha$ line's systemic velocity is allowed to vary independently. On each panel we report the BIC and $\chi^2_r$ for the corresponding model. From these we conclude that there is significant statistical evidence to prefer a model with an offset BLR H$\alpha$ component and no systemic broad emission. 
    }
    \label{fig:niiha}
\end{figure}

\subsection{[\ion{N}{2}]$\,+\,$H$\alpha$ Complex\label{sec:res:g395m}}
We infer a systemic spectroscopic redshift from the narrow H$\alpha$ component to be $z=4.3419$, which is consistent within $1\sigma$ of the redshifts from \cite{Yun2015} measured independently from CO(5-4), CO(4-3), and [\ion{C}{2}]$_{157\,\mu\rm m}$. Hereafter we adopt the narrow H$\alpha$ redshift as our systemic velocity. At first glance, two obvious characteristics stand out: (1) The [\ion{N}{2}]$\,+\,$H$\alpha$ is highly asymmetric with a significant blue wing and sharp red edge, and (2) the [\ion{N}{2}] lines are very strong clearly indicating a powerful AGN. To spectroscopically decompose the [\ion{N}{2}]$\,+\,$H$\alpha$ complex we implement a Markov Chain Monte Carlo fitter using \texttt{emcee} \citep{emcee} that includes flux-conserving resampling to the G395M spectral resolution and accounts for the line spread function. We consider a few models to explore the range of possible profile compositions. In all models we keep all narrow line velocity widths the same, and we fix the [\ion{N}{2}] doublet ratio to $3.047$ \citep{Storey2000}. 

First, we fit a standard broad and narrow line model with systemic velocities fixed to the observed redshift. As expected from the asymmetry of the profile, we find that this model under-predicts the observed blue wing and over-predicts the red edge. We find that we need to allow the systemic [\ion{N}{2}] narrow line velocity to vary in order to reproduce the doublet peaks, and find that the [\ion{N}{2}] lines are kinematically offset from the narrow H$\alpha$ component by $\sim150\,{\rm km\,s^{-1}}$. We speculate that this arises from some amount of host-galaxy H$\alpha$ emission associated with star-forming gas at a different systemic velocity. We next re-fit the data allowing the systemic velocity of the broad H$\alpha$ component to vary freely. This achieves a statistically significant better fit ($\Delta$BIC=$43$) and implies that the broad line component with FWHM$\,=2520\,{\rm km\,s^{-1}}$ is offset from systemic by $-1245\,{\rm km\,s^{-1}}$. We consider the case where both a systemic and offset broad line component are present, but find that this model converges to negligible systemic broad H$\alpha$. The results of these fits are shown in Figure \ref{fig:niiha}. From these fits we conclude that the most likely model is one with an offset broad H$\alpha$ component, which exhibits the lowest $\chi_r^2$ and Bayesian Inference Criterion (BIC). Table \ref{tab:source} includes the derived line properties from this model which we consider fiducial. 

\begin{figure}
    \centering
    \includegraphics[width=\linewidth]{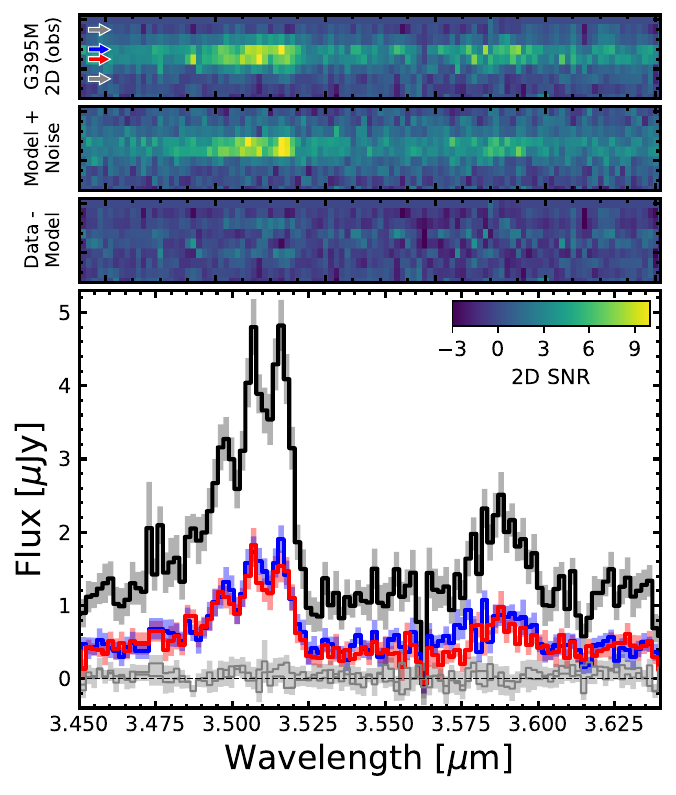}
    \caption{EMBER G395M 2D (\textit{Top panels}) and 1D (\textit{Bottom}) spectrum zoomed in on [\ion{N}{2}]$+\rm H\alpha$ and [\ion{S}{2}]. In the top panels we show the modeled 2D spectrum using \texttt{msafit} and its residuals. The bottom panels compare the extracted 1D spectrum (black, no slit-loss corrections) to single-row extractions (blue, red \-- see corresponding arrows in 2D). We see no differences in the single-row profiles around [\ion{N}{2}]$+\rm H\alpha$, suggesting that both are dominated by the same point source. Grey arrows and corresponding spectra show extractions $\pm2$ pixels away from the trace where the PSF wings are negligable, both of which recover no signal. The modeled 2D spectrum assumes that the broad and narrow lines arise from an unresolved source which successfully reproduces the observed 2D profile. 
    }
    \label{fig:zoom2d}
\end{figure}

\subsubsection{2D G395M Point Source Decomposition}

Different physical origins of the broad, blueshifted component could exhibit different signatures in the 2D G395M spectrum. An AGN broad line region (BLR) would be spatially unresolved and therefore exhibit a point-like trace in the 2D spectrum. Ionized outflows can be extended on spatial scales resolvable by NIRSpec up to $z\sim7$ \citep[e.g.,][]{Liu2026}, with the broad component of H$\alpha$ typically covering $2-3$ kpc in massive star-forming galaxies with similar properties to AzTEC-1 \citep{ForsterSchreiber2014}. To test these scenarios we model the 2D G395M spectrum and look for evidence in the residuals for spatially extended line emission. We use \texttt{msafit} \citep{deGraaff2024} to forward model the NIRSpec G395M PSF at the position in the microshutter array that AzTEC-1's observations were taken. To generate mock 2D spectra we convolve the PSF with a point source and separately with a Sersic profile fit to the F356W morphology. In both cases we account for AzTEC-1's offset source position relative to the shutter center (Fig.~\,\ref{fig:rgb}). We then construct a 2D line emission model using the line fit shown in Figure \ref{fig:niiha}. The emission line component is generated from the point source \texttt{msafit} model convolved with the PSF, and the continuum component from the PSF-convolved Sersic model. The trace center along the spatial axis is assumed to be the same for both components, and is allowed to vary freely at the sub-pixel level as part of the fit. 

Figure \ref{fig:zoom2d} shows a zoom-in on [\ion{N}{2}]+H$\alpha$ and [\ion{S}{2}] in the G395M 1D and 2D, including the 2D source model and its residuals. No residual pixels between $3.45-3.55\,\mu $m, in the vicinity of the broad H$\alpha$ component, have SNR$\,>4$, disfavoring the presence of residual line emission not associated with a point source. As a further test, we also extract 1D spectra from the 2D for the two pixel rows containing most of the signal, and find no significant spatial variation in [\ion{N}{2}]+H$\alpha$. We also find no residual flux beyond 2 pixels of the trace where the wings of the PSF are negligible. Collectively, these tests demonstrate that emission lines in the G395M spectrum are likely to arise from compact source extended on sub-pixel scales $<0.1^{\prime\prime}$ or equivalently $<675$ pc. 

\subsection{[\ion{S}{2}] Doublet\label{sec:doublet}}
We detect the [\ion{S}{2}] doublet in the EMBER G395M spectrum at an integrated SNR of $12.8$, with a peak SNR per channel of $4$. The relatively lower SNR of this complex limits our ability to extract meaningful kinematic and density constraint from a multi-component line decomposition. Nevertheless, we fit the 6716\AA\ and 6731\AA\ doublet with broad and narrow components in similar fashion to [\ion{N}{2}]$\rm + \,H\alpha$, restricting the doublet line ratio in each component to the theoretical range for $T_e=8,000-22,000$ K (Ch.\,5, \citealt{Osterbrock2006}). The only statistically meaningful claim we can make from these fits is that there is very strong evidence ($\Delta \rm BIC>10$) in favor of including a broad [\ion{S}{2}] component with FWHM\,$\sim1000\,{\rm km\,s^{-1}}$ and a blueshifted velocity offset. The inferred kinematic properties are highly uncertain (see Table \ref{tab:source}), and the fits prefer [\ion{S}{2}]6716\AA/[\ion{S}{2}]6731\AA\ $>1$ possibly indicating electron densities $n_e<10^3\,{\rm\,cm^{-3}}$ \citep{Osterbrock2006}; however, [\ion{S}{2}]6716\AA/[\ion{S}{2}]6731\AA\ flux ratios can underestimate the true electron density \citep[e.g.,][]{Holden2026} and ultimately this is degenerate with the velocity widths and offsets. 

\subsection{[\ion{S}{3}] Doublet and Pa$\epsilon$}
We detect the [\ion{S}{3}]$\lambda9068\,\lambda 9545$\AA\ doublet blended with Pa$\epsilon$ in the PRISM spectrum as shown in Figure \ref{fig:spec}. [\ion{S}{3}]$\lambda9068\,\lambda 9545$\AA/[\ion{S}{2}]$\lambda6718,\,\lambda6732$\AA\ ($S_{32}$) is a powerful diagnostic of the ionizing spectrum \citep{Sanders2020}; however, the SNR of each doublet is too low in the G395M spectrum to leverage this diagnostic in light of uncertain broad and narrow components. However, by modeling [\ion{S}{3}]$\lambda9068\,\lambda 9545$\AA\ in the PRISM spectrum we are able to measure a Pa$\epsilon$ flux which constrains the nebular attenuation when combined with H$\alpha$. To fit this complex we assume an intrinsic [\ion{S}{3}]$\lambda9545$\AA/$\lambda 9086$\AA\ ratio of $2.44$ \citep{Mendoza1982,Sanders2020}. During this decomposition we use the continuum model output from our SED modeling which includes Pa$\epsilon$ absorption. We fit a single gaussian component for each feature because PRISM has a resolving power of $R=285$ at $\lambda_{\rm obs}=5\,\mu$m corresponding to $\sim500\,{\rm km\,s^{-1}}$ per pixel. 
From these fits we measure a flux ratio of Pa$\epsilon$/H$\alpha=0.03\pm0.009$. Following the Balmer line ratio method outlined in \cite{Reddy2020} we infer $A_V^{\rm neb}=4.2\pm1.2$. Combinig H$\alpha$ with the non-detection of H$\beta$ yields a limit of $A_V^{\rm neb}>5.3\pm1.6$, consistent with the nebular reddening inferred from Pa$\epsilon$/H$\alpha$ within errors\footnote{We assume $A_\lambda\propto\lambda^{-0.7}$ to go from $A_{\rm H\alpha}$ to $A_V$, consistent with our SED assumptions \citep{Charlot2000}.  $A_V^{\rm neb}$ changes by $<10\%$ assuming a \cite{Calzetti2001} attenuation law, and naturally $A_{\rm H\alpha}$ is determined independent from any dust attenuation law assumption.}. 

\begin{figure}
    \centering
    \includegraphics[width=\linewidth]{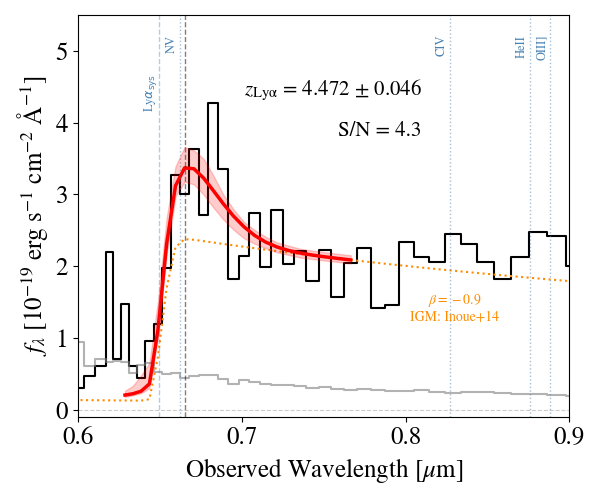}
    \caption{
    Ly$\alpha$ detected in the NIRSpec/PRISM spectrum from EMBER and ZENITH. The orange dashed line shows the modeled UV continuum and Ly$\alpha$ break, using a slope of $\beta=-0.9$ and attenuation by the IGM \citep{Inoue2014}. The dashed blue lines indicate the systemic position of Ly$\alpha$ and other UV emission lines at the spectroscopic redshift measured from narrow lines detected in the G395M spectrum and from ALMA. The red line shows the fitted asymmetric Gaussian profile to the observed Ly$\alpha$ emission line above the modeled continuum (SNR$\,=4.3$). The resulting Ly$\alpha$ redshift corresponds to the wavelength at the peak of the model (vertical black dashed line); however, given that the NIRSpec/PRISM resolution is $R\sim30$ at this wavelength, this offset is unresolved and higher resolution follow-up is required to confirm any true velocity offset.}
    \label{fig:lya}
\end{figure}

\subsection{Ly$\alpha$ Emission Profile}

We detect strong Ly$\alpha$ emission above the modeled continuum with SNR$= 4.3$ in the NIRSpec/PRISM spectrum (Figure \ref{fig:lya}), and a clear Ly$\alpha$ break at 0.648 $\mu$m consistent with the systemic redshift $z_{\rm sys}=4.3419$ from narrow H$\alpha$ and CO. Fitting a linear continuum model we infer an equivalent width of ${\rm EW_{Ly\alpha}=35\pm5}$\AA. The modeled Ly$\alpha$ peak lies at $z_{\rm Ly\alpha} = 4.472$ (3 pixels redward of the systemic redshift), close to what was found from the tentative Ly$\alpha$ detection reported in \cite{Smolcic2011} from a 4 hr integration on Keck DEIMOS. The observed morphology of the line profile is characteristic of high-$z$ Ly$\alpha$ emission lines (a sharp blue cutoff and extended red wing), both suggestive of a large velocity offset due to absorption of the Ly$\alpha$ profile by \ion{H}{1} in AGN outflows and the surrounding ISM \citep{Verhamme2006}. We model the Ly$\alpha$ emission with UV continuum assuming a power law with a slope of $\beta=-0.9$ and a Ly$\alpha$ break set to the systemic redshift with attenuation by the intergalactic medium (IGM) following \citet{Inoue2014}. We then fit an asymmetric Gaussian profile on the emission above the modeled continuum, accounting for the impact of the IGM on the blue-end of the Ly$\alpha$ emission line. These fits are suggestive of scattering through a neutral outflow; However, given that the NIRSpec/PRISM resolution is $R\sim30$ at this wavelength ($\sim9500\,{\rm km\,s^{-1}}$ per pixel, see \citealt{Jakobsen2022}), this offset is unresolved and subject to wavelength calibration uncertainties of even 1 pixel. We therefore do not quote a precise velocity offset and instead caution that higher resolution follow-up with NIRSpec G140M is required to confirm and measure the true Ly$\alpha$ velocity offset. Such observations might also see Ly$\alpha$ in absorption originating from extended star-formation embedded in AzTEC-1's dusty ISM. 

\begin{figure}
    \centering
    \includegraphics[width=\linewidth]{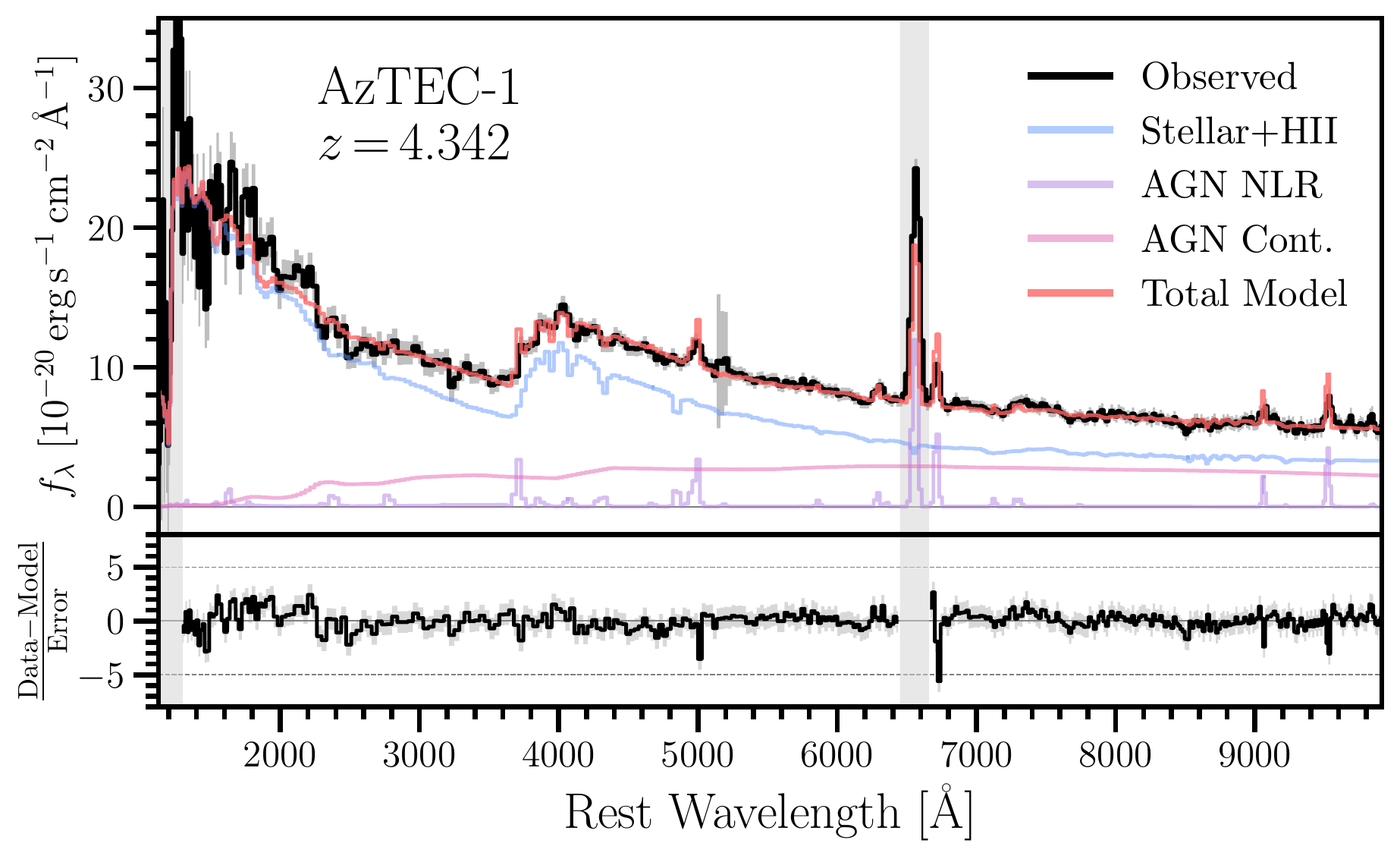}
    \caption{Best-fit \texttt{BEAGLE-AGN} decomposition of AzTEC-1's PRISM spectrum, which is used to infer a stellar mass and the star-formation history. The model includes stellar+\HII\ region emission (blue), NLR emission (purple), and AGN continuum emission (magenta). The H$\alpha+$[\NII] region is masked and these emission line fluxes (measured from the G395M spectrum) are instead directly fitted using the ``spectral indices'' functionality. The best-fit model provides an excellent fit to the observed PRISM continuum, and narrow-component line fluxes are all reproduced within their observational uncertainties. Since \texttt{BEAGLE-AGN} currently does not model broad-line region emission, we exclude broad H$\alpha$ emission from the fit, which is responsible for the visual mismatch between the model and observed H$\alpha$ emission line. The fit prefers modestly attenuated (A$_V$\,$\sim$\,1.5) stellar and AGN continuum emission, with most of the galaxy's stellar mass having formed at a lookback time of $\sim$\,400$-$500 Myr.
    }
    \label{fig:beagleagn}
\end{figure}

\subsection{SED Decomposition and Star-formation history}
We detect a significant Balmer break in the PRISM spectrum, which indicates the presence of an evolved stellar population. The broad H$\alpha$ emission may suggest an unobscured view of an accreting BH, so significant AGN continuum emission may also be present. Thus, to determine the stellar mass and star-formation history (SFH) of AzTEC-1, we perform a spectral decomposition of its observed PRISM spectrum into these various components. To do this, we perform spectro-photometric fitting using a version of the BayEsian Analysis of GaLaxy sEds \citep[\texttt{BEAGLE};][]{Chevallard2016} SED fitting code called \texttt{BEAGLE-AGN} \citep{beagleAGN}. In addition to standard models of stellar \citep{bruzual_stellar_2003} and nebular emission from \HII\ regions \citep{gutkin_modelling_2016}, \texttt{BEAGLE-AGN} self-consistently models emission from the AGN narrow-line region (NLR) using the CLOUDY photoionization grids of \citet{feltre_nuclear_2016}, with updates as described in \citet{mignoli_obscured_2019}. Moreover, an analytic AGN continuum emission model can be included in the fitting.

We generally follow the fitting procedure of \citet{gupta_rapid_2026}, who fit PRISM spectra of broad-line AGN and investigate the impacts of modeling choices on derived parameter estimates. Since \texttt{BEAGLE-AGN} does not currently include models of broad emission, we mask the H$\alpha +$[\NII] region. We instead directly fit the narrow-component H$\alpha$ and [\NII] fluxes as measured from the G395M spectrum in addition to continuum emission from the PRISM. We use the same set of priors listed in Table 1 of \cite{gupta_rapid_2026}, and refer to their Section 3.3 for a full description of the \texttt{BEAGLE-AGN} fits. In short, we adopt a \cite{Chabrier2003} initial mass function and a delayed-exponential SFH with a flexible 10 Myr burst. A two-component dust attenuation model of \cite{Charlot2000} is selected for the stellar and NLR emission. The AGN continuum emission is included as an analytic power-law $f_{\lambda}$\,=\,$f_{1500}(\lambda/1500$\AA)$^{{-1.7}}$, where $f_{1500}$ is the flux at 1500\AA. The attenuation on this AGN continuum is modeled by an SMC dust law \citep{pei_interstellar_1992} and fitted independently of the stellar and NLR attenuation. We fit the PRISM spectrum with a stellar-only model, a second model with stellar and NLR emission, and a third model with stellar, NLR, and AGN continuum emission. The AGN continuum model is selected as the best fit according to a Bayes factor comparison and provides an excellent match to the observed continuum and line fluxes. The best-fit AGN continuum model from \texttt{BEAGLE-AGN} is shown in Figure \ref{fig:beagleagn}, and its corresponding parameters are listed in Table \ref{tab:source}. From these fits we infer that AzTEC-1 formed 90\% of its mass ($t_{90}$) $\sim140$ Myr prior, with a total stellar mass of $\log\,M_*/M_\odot=10.6\pm0.05$ and super-solar stellar ($\log\,Z_*/Z_\odot=0.12$) and NLR ($\log\,Z_*/Z_\odot=0.38$) metallicities. 

To better characterize of the SFH of AzTEC-1, we also test an equivalent set of SED fits with a non-parametric SFH. We define seven logarithmically-spaced bins between the observed redshift and $z=20$, with the first being fixed to 5 Myr, and carry out the same fits with a stellar-only model, another model with stellar+NLR emission, and a final model with stellar+NLR+AGN continuum emission. We find that the fits without AGN continuum emission do not reproduce the continuum shape nor the line fluxes as well as the previous best-fit model. The non-parametric fit with AGN continuum emission provides a good fit, and suggests that the majority of the stellar mass of AzTEC-1 formed at a lookback time of $\sim100-200$ Myr. However, this model does not provide a significantly improved fit compared to the delayed-exponential SFH version. Moreover, it yields a stellar attenuation of $A_V \sim 0.3$, inconsistent with the available far-IR observations. This is almost certainly a result of overfitting \citep[e.g.][]{magrisc._recovery_2015}, as the non-parametric SFH model with an AGN continuum uses by far the largest number of free parameters. As a result, we select the stellar+NLR+AGN continuum model with a delayed-exponential SFH as our fiducial model for the remainder of this work. 

\subsection{X-ray Upper Limit and the AGN Bolometric Luminosity\label{sec:data:xray}}

AzTEC-1 is not detected in the \textit{Chandra} coverage of COSMOS \citep{Elvis2009,Civano2016}, so we use an X-ray upper-limit to place an upper bound on the AGN bolometric luminosity. Using the Chandra Source Catalog V2.1, we downloaded the full-field combined event, \texttt{arf} and \texttt{rmf} stacks of every pointing that contains the source. These full field combined event files are stacked observation detections event files filtered by the appropriate science energy band and have been reprocessed through \texttt{acis\_process\_events} to apply the latest instrument calibrations and the standard event status and event grade filters. Robust upper limits were estimated via the exposure-corrected $0.5-10$ keV count rates within a 2 arcsec aperture. The $0.5-10$ keV absorbed flux upper limits were determined using the \texttt{CIAO} tools function \texttt{aprates} \citep{CIAO}, and estimated assuming an X-ray power-law slope of $\Gamma = 2$ and galactic absorbing column density (N$_{\mathrm{Gal}}$ = 2.7$\times 10^{20}$ cm$^{-2}$, \citealt{Kalberla2005}). We then apply a bolometric correction to the X-ray flux upper-limits via \cite{Shen2020}, and find for moderate obscuration L$_\mathrm{Bol,\,AGN} < 3.39\times 10^{45}$ erg/s (N$_{\mathrm{H}}\sim10^{22}\,\rm cm^{-2}$), and for heavier obscuration, L$_\mathrm{Bol,\,AGN} < 6.68\times 10^{45}$ erg/s (N$_{\mathrm{H}}\sim10^{24}\,\rm cm^{-2}$). Thus, regardless of obscuration, we find the X-ray derived bolometric luminosity upper-limits consistent with the BEAGLE-AGN decomposition of $\log\,L_{\rm Bol,\,AGN}/{\rm erg\,s^{-1}}=44.9\pm0.1$.

\begin{figure*}
    \centering
    \includegraphics[width=\linewidth]{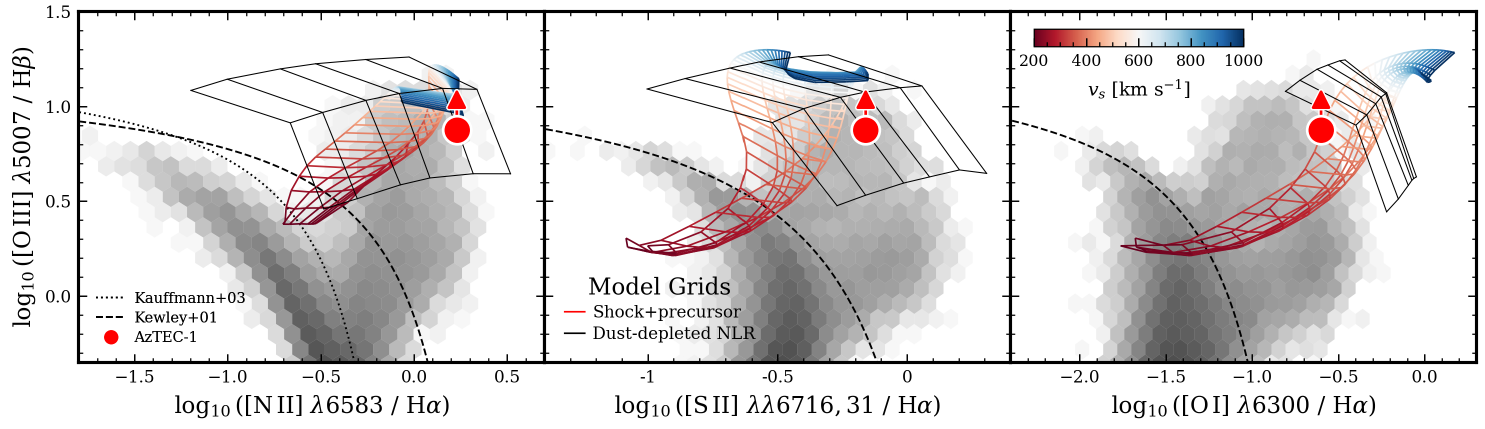}
    \caption{BPT diagrams showing AzTEC-1's narrow emission line ratios (red circle) compared to $z<0.4$ SDSS galaxies (grey) and photoionization models. We show the BPT space covered by a super-solar metallicity ($\log\,Z/Z_\odot=2$) shock from MAPPINGS V \citep{MAPPINGSV} with $v_s\in[200,1000]\,\rm km\,s^{-1}$ and magnetic field strength $B\in[10^{-4},10]\,\mu$G, including a photoionized precursor with a preshock density of $n=1\,\rm cm^{-3}$. In black we show the dust-depleted AGN NLR models from \cite{Zhu2023} with $\log U\in[-3.4,-2]$ and $Z/Z_\odot\in[0.5,2.5]$. AzTEC-1 is within the general AGN BPT domain \citep{Kewley2001,Kauffmann2003}, exhibiting some overlap with both the high velocity shock and the dust-depleted NLR models. 
    }
    \label{fig:bpt}
\end{figure*}

\subsection{Morphology}

AzTEC-1 exhibits a compact, smooth morphology in the NIRCam imaging with no obvious evidence for a recent major merger in the form of tidal tails or multiple components. We measure non-parametric morphological parameters in each NIRCam band including the concentration (C) and asymmetry (A) \citep{Conselice1997,Conselice2003}, and find $C\in[2.5-4]$ and $A\in[0.1-0.2]$. While these indicators are less capable at separating morphological classes in \textit{JWST} observations of $z=3-9$ galaxies \citep{Kartaltepe2023}, they do support the general interpretation that AzTEC-1 is compact and smooth. 

NIRCam F115W and F150W probe the rest-UV whereas F277W overlaps AzTEC-1's Balmer Break with minimal line contamination. The resolution of NIRCam at these wavelengths is $0.04-0.09^{\prime\prime}$, which corresponds to a physical limit of $\sim0.27$ kpc in the rest-UV and $0.62$ kpc in the rest-frame optical. AzTEC-1 has a UV half-light radius of $0.37\pm0.02$ kpc and an optical half-light radius of $0.72\pm0.06$ kpc in the COSMOS-Web morphological catalog from \cite{Yang2026}. These half-light radii are more compact than the typical massive radio-selected SFG at $z\sim2-3$ \citep{JimenezAndrade2021} and similar \textit{JWST}-selected galaxies at $z=3-9$ \citep{Miller2025,Allen2025}. 

We compute flux-weighted centroids in each NIRCam band relative to the resolved and unresolved ALMA data. We find an average offset between the NIRCam imaging and the unresolved ALMA position of $0.040^{\prime\prime}$ ($270$ pc), which confirms the offset of $0.2^{\prime\prime}$ between ALMA and \textit{HST} imaging reported by \cite{Iono2016} who discuss evidence for $3-4$ kpc extended dust emission. The NIRCam centroids from the different bands agree with the HST center and each-other within the accuracy of the PSF, and overlap with the resolved ALMA dust continuum as shown on Figure \ref{fig:rgb}.  
\section{Discussion\label{sec:disc}}

As the brightest mm-selected galaxy in COSMOS AzTEC-1 is relatively extreme with regards to its high total IR luminosity, dust mass, redshift, and stellar mass \citep{Smolcic2011,Yun2015}. 
AzTEC-1's extreme properties extend to its rest-frame optical emission line kinematics detected in G395M. For context, high-redshift AGN with broad, offset velocity components tend to have offset Balmer line velocities less than what we find in AzTEC-1. \cite{Maiolino2024} report candidate dual AGN at $z\sim4-6$ with $\log\,M_*/M_\odot\sim8-10$ and velocity shifts of $\sim100-400\,{\rm km\,s^{-1}}$. Similar kinematic offsets $<500\,{\rm km\,s^{-1}}$ are seen in a $z\sim7$ quasar with H$\alpha$ \citep{Bosman2024}, and \cite{Ubler2023} report broad H$\alpha$, H$\beta$, and [\ion{O}{3}] components with FWHM$\sim1000-3000\,{\rm km\,s^{-1}}$ offset from the narrow line systemic velocity by only $\sim100\,{\rm km\,s^{-1}}$. 
Broad and blue-shifted [\ion{O}{3}] components with kinematic offsets and widths similar to AzTEC-1's H$\alpha$ component appear to be common in $z>5$ quasars \citep{Yang2023,Liu2026}, which are commonly interpreted as evidence for outflows but these kinematic signatures are also components of selecting candidate binary or runaway SMBHs \citep{Barrows2025,Lu2025}. 

Figure \ref{fig:bpt} shows where AzTEC-1 falls along Baldwin–Phillips–
Terlevich (BPT) emission line diagnostic diagrams \citep{Baldwin1981}. Shock model grids from MAPPINGS V \citep{Allen2008,MAPPINGSV} were obtained from the Mexican Million Models database (3MdB, \citealt{3MdB}), and we also compare to the dust-depleted AGN NLR models of \cite{Zhu2023}. We measure dust-corrected [\ion{N}{2}]/H$\alpha=1.7$ and [\ion{O}{3}]/H$\beta>7.5$ using just narrow line components, supporting the presence of an AGN \citep{Baldwin1981,Veilleux1987,Kewley2001,Kauffmann2003}. We also find [\ion{O}{1}]/H$\alpha=0.25$ and [\ion{S}{2}]/H$\alpha=0.7$, consistent with dust-depleted NLR and $n_e\sim150-1000\,\rm cm^{-3}$ that is allowed by the uncertainty in the observed [\ion{S}{2}] doublet ratio \citep{Zhu2023}. As shown in Figure \ref{fig:bpt}, these optical emission line ratios are also reproducible by models of radiative shocks that include a photoionized precursor \citep{Allen2008,MAPPINGSV}, which could be powered by an outflow; however, AzTEC-1 is generally preferred by the AGN NLR grids, and the shock velocities needed to recreate its narrow line ratios along each BPT diagram are $v_s>400\,\rm km\,s^{-1}$, and as high as $1000\,\rm km\,s^{-1}$, in excess of the narrow line widths we measure with $290^{+7}_{-16}\,\rm km\,s^{-1}$.

We now discuss the likelihood that AzTEC-1 hosts an outflow, a candidate binary SMBH system, or a recoiling AGN. We begin with discussing the outflow scenario as this is the most likely as a matter of precedent. Ultimately, an outflow is disfavored on the basis of several arguments that make use of \textit{JWST} and ALMA spectroscopic and imaging data. We conclude with discussing the binary and recoiling SMBH, of which the former is preferred largely by timescale arguments. 

\subsection{Outflows}
The shape of AzTEC-1's Ly$\alpha$ profile (Fig.\,\ref{fig:lya}) is suggestive of a neutral outflow, which might indicate that the blueshifted broad H$\alpha$ traces the ionized component. 
As discussed in Section \ref{sec:doublet}, the [\ion{S}{2}] doublet detected in G395M also hosts a broad and blueshifted component with a flux ratio of [\ion{S}{2}]6716\AA/[\ion{S}{2}]6731\AA$\,=1.12^{+0.27}_{-0.37}$ suggestive of $n_e\approx500\,{\rm cm^{-3}}$ \citep{Osterbrock2006}, in general agreement with electron densities measured in ionized outflows at high-redshift \citep{Harrison2014,Vayner2017,ForsterSchreiber2019}, but with significant uncertainties. 

\begin{figure}
    \centering
    \includegraphics[width=\linewidth]{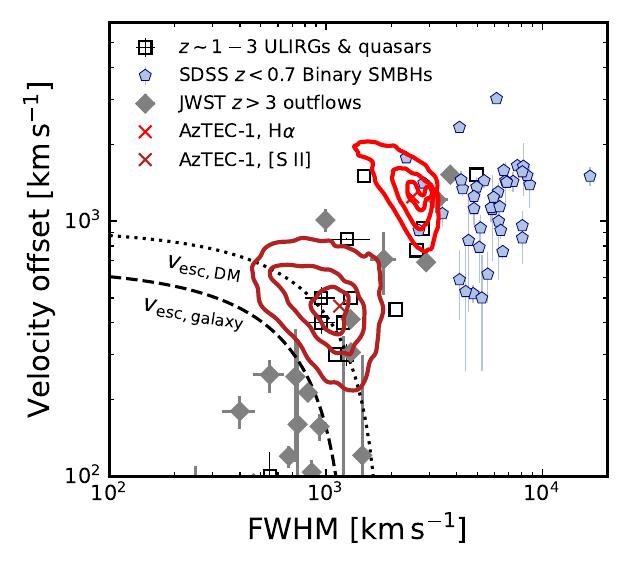}
    \caption{Velocity offsets ($|\Delta v|$) and broad component FWHM from rest-frame optical emission lines. Outflows at low-redshift (black squares, \citealt{Harrison2012,Zakamska2016}) and high-redshift (grey diamonds, \citealt{Carniani2024,Xu2025,Yang2023,Liu2026}) are typically measured from [\ion{O}{3}]. Binary SMBH candidates at $z<0.7$ (blue pentagons, \citealt{Eracleous2012}) are inferred from H$\beta$. AzTEC-1's broad H$\alpha$ (red) and broad [\ion{S}{2}] (orange) are shown with contours spanning the 16th, 50th, and 84th percentiles of all models considered with an offset broad component. The measured velocities exceed the escape velocity of AzTEC-1's galaxy ($10^{10.5}\,M_\odot$ at 1 kpc, dashed) and dark matter halo ($10^{12}\,M_\odot$ at 10 kpc, dotted) above which the maximum gas velocity ($|\Delta v|+0.5\times$FWHM) would be unbound. All other outflows with $|\Delta v|$ and FWHM exceeding $1000{\rm\,km\,s^{-1}}$ are extended over $2-15$ kpc \citep{Harrison2012}. AzTEC-1's broad H$\alpha$ kinematics overlap the most extreme outflows in luminous $z\sim5-6$ quasars and SMBH binary candidates selected from SDSS.
    }
    \label{fig:outflows}
\end{figure}

The velocity offset and width of the observed broad H$\alpha$ and [\ion{S}{2}] components are both larger than what cosmological zoom-in simulations reproduce \citep{Nelson2019,Kostyuk2025}. Notably, both the velocity offset and width of the broad H$\alpha$ emission is $\sim2\times$ that of what is found in [\ion{S}{2}]. These are found to agree with one-another in local AGN driven outflows \citep{KovaceviDojcinovic2022}, although we note that stratification of a multiphase outflow might also play a role \citep{Revalski2021,Holden2026}.  As shown in Figure \ref{fig:outflows}, the kinematics inferred from H$\alpha$ and [\ion{S}{2}] are on the high end for what has been reported in ULIRGs, radio galaxies, $z\sim1-3$ quasars and some $z>5$ quasars \citep{Harrison2012,Zakamska2016,Yang2023,SaldanaLopez2025,Liu2026}. We include for reference lower mass $z\sim3-7$ galaxies from \textit{JWST} \citep{Carniani2024,Xu2025}. 
A subset of the $z\sim1-2$ ULIRGs and $z>5$ quasars reach outflow velocity offsets and broad FWHM exceeding $1000\,{\rm km\,s^{-1}}$ like AzTEC-1, and these outflows are extended over many kpc \citep{Harrison2012,Liu2026}. Thus we might expect a similar spatial extent in AzTEC-1, also by virtue of the AGN-driven multiphase outflow characteristics recovered in simulations \citep{Nelson2019}. As discussed in Section \ref{sec:res:g395m} we find no evidence for extended line emission on scales greater than $\approx600$ pc. 

\begin{figure}
    \centering
    \includegraphics[width=\linewidth]{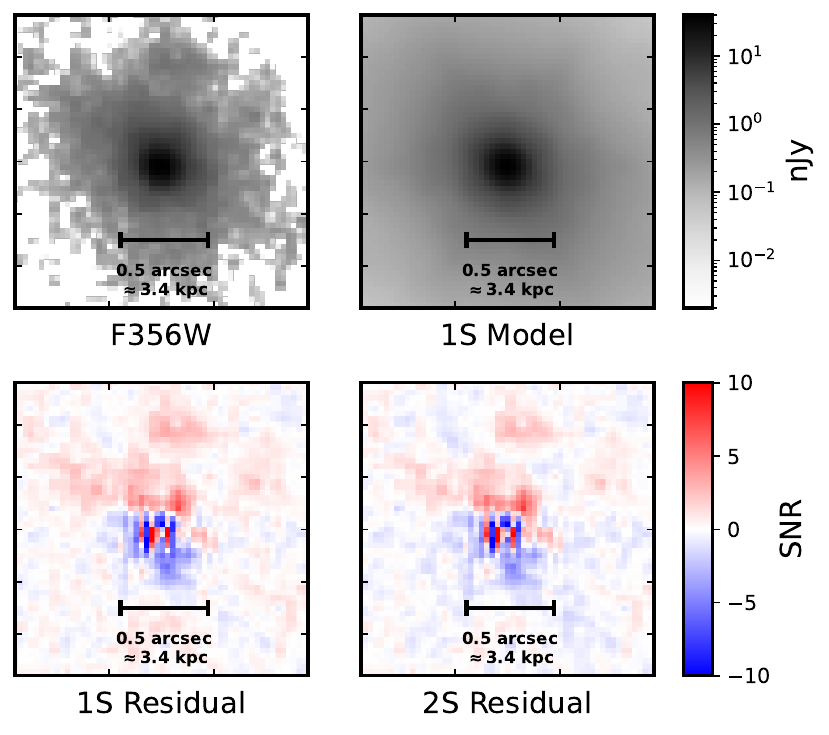}
    \caption{S{\' e}rsic decomposition of AzTEC-1's F356W image (\textit{Top left}) which overlaps [\ion{N}{2}]+H$\alpha$. These emission lines contribute $\approx10\%$ to the total flux. We show the model from a single S{\' e}rsic fit (1S, \textit{Top right}) which looks identical to a double S{\' e}rsic fit (2S). Residual maps are shown in units of SNR (\textit{Bottom}). Both S{\' e}rsic models converge to a point-source dominated profile $n_{\rm Sersic}\gg10$. There is some evidence for residual asymmetric structure in the model-subtracted maps within the inner few kpc that could be attributable to an ionized outflow; however, the aggregate flux of this residual emission does not reproduce the total flux from broad H$\alpha$.}
    \label{fig:f356wmorph}
\end{figure}

To test for morphological evidence of an extended outflow in the NIRCam imaging, we fit and subtract a S{\' e}rsic model to the F356W image using \texttt{statmorph} \citep{RodriguezGomez2019} and an empirical F356W PSF constructed from the COSMOS imaging data (\citealt{Franco2026}, Akins et al., in prep.). We expect $\approx0.15\,\mu$Jy of extended broad H$\alpha$ emission in the F356W image if it were a kpc-scale outflow. The residual image shown in Figure \ref{fig:f356wmorph} recovers only $\sim0.08\,\mu$Jy in total, only half of what is expected from the broad H$\alpha$ component alone disregarding the equal amount of flux expected from the narrow [\ion{N}{2}]+H$\alpha$ lines. This is consistent with no extended emission in the F356W data. 

\begin{figure}
    \centering
    \includegraphics[width=\linewidth]{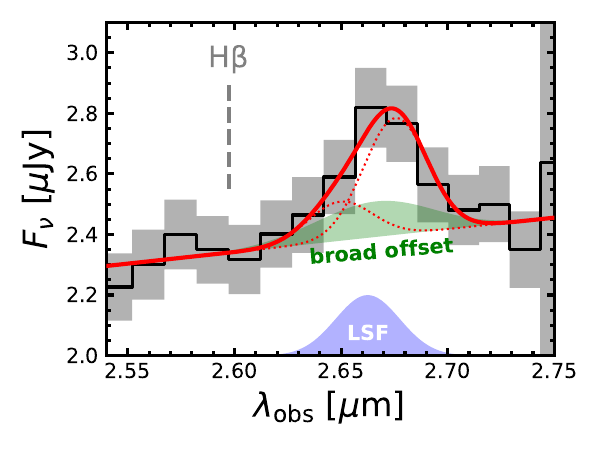}
    \caption{NIRSpec/PRISM zoom-in on the [\ion{O}{3}] doublet and non-detection of H$\beta$. The red line shows a model fitting only narrow [\ion{O}{3}], but this fit is dominated by the PRISM line spread function (blue). The expected offset broad component for [\ion{O}{3}] based on the G395M H$\alpha$ profile is shown as a shaded green region, which can be added to the model but with no discriminatory power given the resolution.}
    \label{fig:oiiihb}
\end{figure}

An ionized outflow detected in H$\alpha$ should also appear in [\ion{O}{3}] \citep[e.g.,][]{Ubler2023,Vayner2025}, but PRISM has a resolving power of $R\sim75$ at $\lambda\sim2.6\,\mu$m corresponding to a line spread function with a width of $\sim2500\,{\rm km\,s^{-1}}$. Figure \ref{fig:oiiihb} shows the decomposition of the [\ion{O}{3}] doublet which is fully explained by just narrow components; however, adding the offset broad velocity profile expected from what is measured in H$\alpha$ can achieve the same goodness-of-fit. Therefore we cannot  definitively rule in favor of or against an ionized outflow from the marginally resolved [\ion{O}{3}] doublet in the PRISM data.

High angular resolution ($<500$ pc) ALMA observations of dust and multiphase emission line diagnostics provide some evidence against a outflow. \cite{Tadaki2019} measure [\ion{C}{2}]$_{157\,\mu \rm m}$ and [\ion{N}{2}]$_{205\,\mu \rm m}$ in the inner kpc of AzTEC-1 with a resolution of $0.3^{\prime\prime}$. \cite{Tadaki2018} resolve the core in CO(4-3) at $0.03^{\prime\prime}$ resolution. These studies find consistent kinematic properties between the different tracers with $v_{\rm max}\approx230\pm20\,{\rm km\,s^{-1}}$, revealing no evidence for a high velocity outflow component across the neutral, molecular, and warm ionized phases that these emission lines can be excited in despite the fact that stratification is expected from the multiphase structure of the outflow \citep[e.g.,][]{Richings2021}. Indeed, \cite{Vayner2021} observe kpc-scale outflows in H$\alpha$, [\ion{O}{3}], and CO in $z\sim1-2$ quasars. \cite{Tadaki2020} do find evidence for non-corotating gas in residual [\ion{C}{2}] emission in AzTEC-1's inner few hundred pc after subtracting out the rotationally-dominated emission, but these residual clumps have $\leq200\,{\rm km\,s^{-1}}$ kinematic offsets from systemic. The velocity dispersion \cite{Tadaki2020} measure in AzTEC-1's inner few hundred pc is $110\,{\rm km\,s^{-1}}$, with a maximum rotational velocity of $<75\,{\rm km\,s^{-1}}$.  Thus there is no evidence from these high-resolution ALMA studies for outflowing ionized gas with velocities in excess of $\approx200\,{\rm km\,s^{-1}}$. As a caveat, we note that ionized gas outflows in nearby galaxies hosting kpc-scale radio jets sometimes do not show up in cold/warm molecular gas \citep{RamosAlmeida2017,Runnoe2018,RamosAlmeida2022}. More generally, it is unclear how efficiently an AGN-driven ionized gas outflow couples to the dense molecular gas, as this likely depends on the gas geometry, molecular-cloud covering fraction, cooling and dissociation processes, and the age of the outflow \citep{Richings2018,Veilleux2020}. The ionized wind could instead escape through lower-density channels without entraining or accelerating enough molecular gas to produce a detectable neutral/molecular outflow.

Should the Ly$\alpha$ offset be confirmed, this could mean that the neutral gas scattering Ly$\alpha$ photons is extended on much larger scales possibly driven by recent star-formation or past episodes of AGN activity. This would be consistent with the fact that the low and high angular resolution ALMA data have different centroids, indicative of resolving out large scale emission from dust in the ISM of the host galaxy. 
The outflow could be relatively metal-poor, but this is inconsistent with the evidence for a broad component in [\ion{S}{2}], as well as the significant metal enrichment observed in the central kpc \citep{Tadaki2019} where presumably such an outflow would be launched. Alternatively, these lines might be suppressed in an ionized outflow due to density effects. [\ion{C}{2}]$_{157\,\mu m}$ and [\ion{N}{2}]$_{205\,\mu m}$ have critical densities of $\sim50\,\rm cm^{-3}$ for collisions with free electrons, while [\ion{O}{3}]$_{88\,\mu m}$ has a critical density of $500\,\rm cm^{-3}$. On the other hand, $n_{\rm crit}\approx10^3\,\rm cm^{-3}$ for [\ion{C}{2}]$_{157\,\mu m}$ collisionally excited by Hydrogen and can therefore trace neutral and molecular phases of an outflow \citep[e.g.,][]{Janssen2016,Parlanti2025}. 

As shown on Figure \ref{fig:outflows}, the putative velocities from H$\alpha$ and [\ion{S}{2}] exceed the escape velocity of AzTEC-1 and its dark mater halo\footnote{We use the total stellar mass and optical size of $\approx1$ kpc to infer the galaxy's escape velocity, and assume a Dark Matter mass of $10^{12}\,M_\odot$ from \cite{Behroozi2013}.} and should plausibly have some extended morphological signature. The only other galaxies with similar kinematics all exhibit kpc scale outflows \citep{Harrison2012,Liu2026}, which in principle could be absent from AzTEC-1 only if the outflows were launched relatively recently ($<10^5$ yr) to be unresolved by \textit{JWST} and leave no imprint on the gas properties measured by ALMA within the inner $300$ pc.  

If the outflow is spatially unresolved it should be extended over $R<675$ pc based on our decomposition of the 2D G395M spectrum (Fig.\,\ref{fig:zoom2d}). We estimate a lower limit on the outflow mass loss rate following \cite{Osterbrock2006} and \cite{Vayner2025wise} assuming $n_e=500\,{\rm cm^{-3}}$, and $L_{\rm H\alpha,\,broad}=1.6\times10^{10}\,L_\odot$ after correcting for dust attenuation. For the outflow velocity we adopt $v_{\rm outflow}=v_{\rm 10 ,\,H\alpha}=4500\,{\rm km\,s^{-1}}$, equal to where 10\% of the line is integrated given that the line wings represent the true velocity of most of the outflowing material \citep{CanoDiaz2012,Greene2012,Vayner2025wise}. This works out to an ionized outflow mass of $\log\,M_{\rm ion}/M_\odot\approx8.7$ and a corresponding outflow rate of at least $3700\,M_\odot\,\rm yr^{-1}$. 
This is greater than the maximum reported outflow rate (derived from broad H$\alpha$) of $\approx2500\,{M_\odot\,\rm yr^{-1}}$ reported in high-redshift quasars \citep{Vayner2021c,Vayner2025wise}. 
For context, the most luminous obscured quasar in the Universe with $\log\,M_*/M_\odot=11.6$ and SFR$\,\approx500\,M_\odot\,\rm yr^{-1}$ hosts a nuclear outflow with  $\log\,M_{\rm ion}/M_\odot=8.3\pm0.3$ and $\dot M_{\rm ion}=2000\pm800\,M_\odot\,\rm yr^{-1}$ \citep{DiazSantos2018,Vayner2025wise}, embedded within H$\alpha$ emission extended over 10s of kpc with radial velocities $\in[-1000,1000]\,{\rm km\,s^{-1}}$. 
With $\log\,L_{\rm Bol,\,AGN}/{\rm erg\,s^{-1}}=44.9\pm0.1$ (Sec.\,\ref{sec:data:xray}), we find that the ratio of the outflow's momentum flux ($\dot P_{\rm outflow}=\dot M_{\rm outflow}\times v_{\rm outflow}$) to that of the accretion disk ($\dot P_{\rm AGN}=L_{\rm Bol,\,AGN}/c$) is $\dot P_{\rm outflow}/\dot P_{\rm AGN}\approx2-4\times10^3$, at least an order of magnitude greater than that of $z=2-3$ red type-2 quasars \citep{Vayner2021c}.  
Sustained at its current rate, AzTEC-1's nuclear outflow rate would rapidly deplete ($<1$ Myr) an amount of gas equivalent to the molecular reservoir measured in its inner kpc, and ultimately make AzTEC-1 host of one of the most extreme AGN-driven outflows observed to-date if this scenario is confirmed.

It is unlikely that such an energetic outflow would leave no kinematic imprint on the multiphase gas traced by the deep and spatially resolved multi-wavelength observations in AzTEC-1 from \textit{JWST} as well as ALMA. While future observations are needed to conclusively rule out the presence of an outflow in AzTEC-1, we consider this scenario disfavored by the arguments summarized below: 
\begin{enumerate}
    \item There is no evidence for a spatially extended outflow component in the 2D NIRSpec spectrum, NIRCam imaging, and ALMA spectrophotometric data on scales above a few hundred pc. 
    \item ALMA observations of multiphase gas diagnostics in the inner 300 pc find maximum velocities of $230\,{\rm km\,s^{-1}}$ \citep[e.g.,][]{Tadaki2018,Tadaki2019}. 
    \item The energetics of a putative outflow would be greater than the most extreme outflows observed to date in $z\sim2-5$ quasars, all of which host extended (kpc-scale) and multiphase components \citep{Vayner2021,Vayner2021c,Vayner2025wise} that are not found in AzTEC-1.  
\end{enumerate}

\subsection{Supermassive Binary Black Holes}

We now consider the scenario whereby the observed broad H$\alpha$ velocity offset arises from a SMBH binary with one active BLR, as in \cite{Bogdanovic2009}, who present a model to explain a $z=0.71$ quasar with an offset BLR velocity of $2650\,{\rm km\,s^{-1}}$ \citep{Komossa2008}. The separation between both black holes is sub-pc, and the lower mass orbiting companion host the BLR that preferentially accretes gas from the NLR starving the more massive primary SMBH and giving rise to the broad, blueshifted H$\alpha$ emission. Within this framework the broad and blueshifted [\ion{S}{2}] is produced in accretion streams feeding the secondary from the inner edge of the circumbinary disk, which \cite{Bogdanovic2009} predict to be lower in velocity width (as seen in AzTEC-1) shifted to approximately the velocity of the BLR, but with some variability over the orbital cycle that might explain the $\Delta v_{\rm [S\,II]}<\Delta v_{\rm H\alpha}$ we observe.   
In this model the BLR sets the mass of the companion and therefore a lower limit on the total mass of the binary system. 

\begin{figure}
    \centering
    \includegraphics[width=\linewidth]{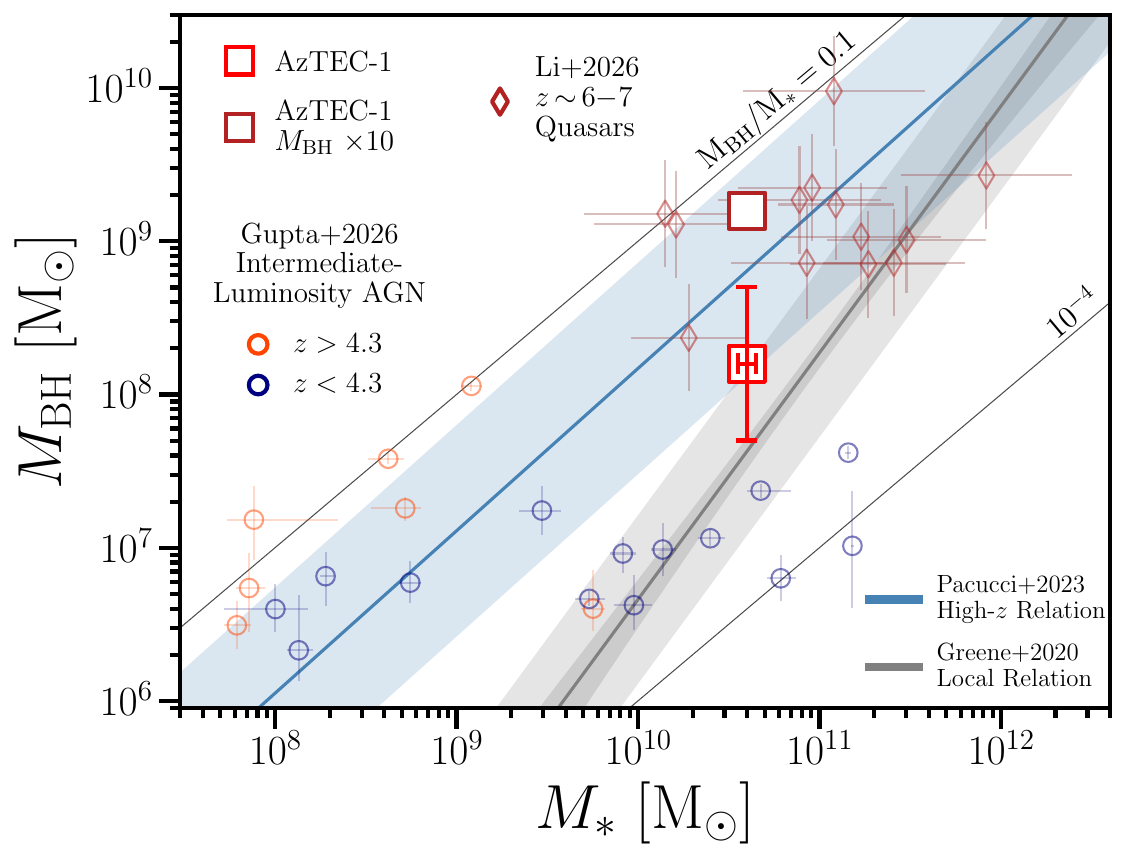}
    \caption{Measured black hole mass to stellar mass ratio for AzTEC-1's companion SMBH (red square) compared to the local relation \citep[grey shaded region;][]{greene_intermediatemass_2020} and \textit{JWST}-observed AGN at $z\sim4-7$ \citep[blue shaded region;][]{Pacucci2023}. If the active BH is in a near-equal mass binary, then AzTEC-1 would be consistent with the local relation and may represent a source which previously had an elevated $M_{\rm BH}/M_{\rm *}$ ratio but whose host galaxy rapidly assembled in the past $\sim500$ Myr. This transition may be echoed by the elevated $M_{\rm BH}/M_{\rm *}$ ratios of \textit{JWST} AGN above the redshift of AzTEC-1 (orange circles)  compared to those at lower redshifts (blue circles). The maroon square represents the total binary mass in AzTEC-1 assuming a SMBH binary mass ratio of $q=0.1$; in this case, the $M_{\rm BH}/M_{\rm *}$ ratio remains high, near $z>6$ quasars \citep[maroon diamonds;][]{li_dichotomy_2026}. It may be challenging to reconcile such a system with the local relation given the limited available gas budget \citep{Tadaki2018}, though such extreme sources do exist in the low-redshift universe \citep{mezcua_overmassive_2024}.
    }
    \label{fig:mbhmstar}
\end{figure}

Using the H$\alpha$ calibration of \cite{reines_dwarf_2013} and our dust-corrected broad H$\alpha$ luminosity, we estimate a mass of the companion SMBH hosting the BLR to be $\log M_{\rm BH}/M_\odot=8.2\pm0.5$. Figure \ref{fig:mbhmstar} shows the mass of AzTEC-1's companion relative to the local relation between SMBH mass and host galaxy stellar mass \citep{greene_intermediatemass_2020} and an empirical fit to $z \sim 4-7$ \textit{JWST}-observed AGN \citep{Pacucci2023}. For context, we also show the position of AzTEC-1 with $10\times M_{\rm BH}$ to represent the total mass of a binary system with a mass ratio $q=0.1$. 

If AzTEC-1 hosts an equal mass SMBH binary, then it would lie within the intrinsic scatter of both the local and high$-z$ $M_{\rm BH}-M_*$ relation. Interestingly, \citet{gupta_rapid_2026} find that AGN at $z > 4.5$ appear to display elevated $M_{\rm BH}/M_{\rm *}$ ratios, whereas sources at $z < 3.5$ are consistent with the local relation. This rapid evolution over a period of just $\sim 500$ Myr might be facilitated by merger-enhanced star formation, a phenomenon observed in certain theoretical models \citep[e.g.][]{trinca_episodic_2024}. The SFH of AzTEC-1 suggests that it may have just completed such a transition; its $M_{\rm BH}/M_{\rm *}$ ratio may have been elevated at a lookback time of  $>500$ Myr, consistent with \textit{JWST} high-redshift AGN, but a merger-triggered burst of star formation may have brought it to the local relation.

On the other hand, if the mass ratio of the SMBH binary is $q\leq0.1$, then AzTEC-1 has a $M_{\rm BH}/M_{\rm *}$ ratio enhanced by $\gtrsim 1$ dex compared to the local relation. This is a region largely occupied by UV-bright quasars at $z \gtrsim 6$ \citep{li_dichotomy_2026}; AzTEC-1 may thus be a descendant of one of these sources. 
However, it may be challenging for such a system to then reach the local relation, given that its [\ion{C}{1}]-based gas mass of $\sim 7\times10^{10}$ \citep{Tadaki2018} would have to be converted into stars extremely efficiently even assuming no further BH mass growth. AzTEC-1 may thus be a progenitor of rare low-redshift sources which display elevated $M_{\rm BH}/M_{\rm *}$ ratios \citep[e.g.][]{mezcua_overmassive_2024}, or perhaps may resemble the high-redshift quasar in \citet{stone_z708_2025}, only eventually reaching the local relation assuming substantial accretion from the IGM or through additional major mergers.

The putative binary separation and period depend on the orbital inclination $i$.
\cite{Tadaki2018} constrain the inclination of the nuclear gas disk to $i=44\pm1^{\circ}$ from spatially-resolved CO(4-3) kinematics. Assuming that the sub-pc binary is coplanar with the disk, an equal-mass binary ($q=1$) would have a separation of $a\approx0.05$ pc and an orbital period of $P=65$ years while a $q=0.1$ binary ($10^{8.2}\,M_\odot$ companion orbiting a $10^{9.2}\,M_\odot$ primary) has $a\approx1$ pc and $P\approx2000$ years. Assuming random orientations yields similar values (0.08 pc / 120 yr and 1.5 pc / 4000 yr respectively). In both inclination scenarios the equal-mass binary is well within the final parsec, while a $q = 0.1$ binary would sit near or just outside $1$ pc. The binary plane need not 
align with the host galaxy's disk, and a near-face-on geometry would harden the $q = 0.1$ orbit to 0.17 pc.
The putative binary SMBH orbit is likely close to or within the final pc which is thought to last $\approx10$ Myr \citep{Khan2016,Sobolenko2021}.  

Assuming local scaling relations between H$\alpha$ BLR sizes and luminosity \citep{Mandal2021}, which may not be applicable to binary systems but provide a first order test, we estimate the size of the BLR to be $R_{\rm BLR}=7\pm3\times10^{-3}$ pc which is within the secondary's Roche lobe ($\approx 0.21a$ for $a=1$ pc and $q = 0.1$; \citealt{Eggleton1983}), in which case the BLR would be dynamically bound to the secondary. For an equal-mass binary at the smallest separations the BLR instead approaches its Roche limit and could be tidally truncated. We note that $R_{\rm BLR}/c=8.4$ days so no BLR variability would be expected over the 2 day rest-frame time delay separating the EMBER and ZENITH PRISM observations. 

Observing a change in the broad H$\alpha$ component's velocity offset would provide a direct test of the binary scenario as the BLR orbits the binary center of mass. Assuming $i=44^{\circ}$ the velocity offset would change by $\approx1\,{\rm km\,s^{-1}\,yr^{-1}}$ for $q=0.1$ and  $\approx23\,{\rm km\,s^{-1}\,yr^{-1}}$ for $q=1$ \citep[e.g.,][]{Bogdanovic2009}. These change to $9\,{\rm km\,s^{-1}\,yr^{-1}}$ and $300\,{\rm km\,s^{-1}\,yr^{-1}}$ respectively if the system is more face on ($i<17^{\circ}$, \citealt{Yun2015}). 
Future NIRSpec G395M follow-up could therefore test the equal-mass SMBH binary scenario by measuring a change in the BLR offset over a timescale of a few years, but longer baseline monitoring is needed to be sensitive to velocity changes in an edge-on $q=0.1$ binary. 

SMBH binary infall stalls at orbital radii below $\approx1$ pc in the absence of scattering material due to inefficient energy transport through gravitational waves \citep{Begelman1980,Milosavljevic2001,Merritt2005,Aarseth2008}.  
Cosmological zoom-in simulations predict that the loss cone\footnote{The position and velocity space occupied by stars that will be ejected by the binary SMBH system.} is efficiently re-filled in massive galaxies with surface densities well-matched to AzTEC-1 \citep{Khan2016,Khan2018}, leading to short delays times of $\mathcal{O}(10\,\rm Myr)$.
AzTEC-1's star-formation history suggests that most of the stellar mass formed $\sim100-400$ Myr ago. 
If this past star-formation was merger induced, this would imply a longer delay time between AzTEC-1's galaxy-galaxy merger and SMBH coalescence ($\tau_{\rm delay}$) on the order of a few hundred Myr. 
Timescales of this magnitude can be realized in gas-rich nuclei, specifically via scattering with $\approx10^6\,M_\odot$ gas clumps \citep{Fiacconi2013,Roskar2015,SouzaLima2017}, which would naturally arise in AzTEC-1's dynamically unstable disk \citep{Tadaki2018}. 
\cite{Iono2016} report an integrated flux of $1.54\pm0.15$ mJy within the central $300$ pc of AzTEC-1. For a range in dust temperatures between $25-50\,$K, this implies $\approx10^8\,M_\odot$ worth of dust and a total gas mass of $\sim10^9-10^{10}\,M_\odot$ depending on the assumed dust-to-gas ratio \citep{Clark2023}, consistent with the gas mass estimate in the inner $\sim500$ pc from \cite{Tadaki2018} using CO(4-3) and in excess of the SMBH mass we estimate. Around $\sim20\%$ of this mass is contained in the brightest unresolved nuclear clump, across $\approx100$ pc. There is significant gaseous material in AzTEC-1's nucleus that might have stalled the evolution of the binary at larger orbital radii, and we may be now witnessing the final sub-pc stage after that delay.

Future gravitational wave interferometers like LISA will be capable of detecting the gravitational wave signatures from individual SMBH merger events \citep{LISA2025}. 
While $z>4$ binaries as massive as what AzTEC-1 might host are below LISA's low-frequency cutoff, this system does provide a window into the mechanisms powering low frequency gravitational waves. 
In particular, the delay time between galaxy-galaxy and SMBH mergers is one of the dominant uncertainties when predicting LISA's high-mass, high-redshift merger population \citep{Klein2016}. 
Mock LISA catalogs with $\tau_{\rm delay}$\footnote{SMBH mergers with total mass $\sim10^8$ in  \cite{Singh2026} reach $\tau_{\rm delay}\approx0.1$ Gyr, whereas their distribution in $\tau_{\rm delay}$ for all systems is centered on a Hubble time.} comparable to what we infer 
reach the mass range of AzTEC-1 albeit at $z\approx3$ \citep{Fang2025,Singh2026}, and predict higher occurrence rates for LISA at $z>3$. 
Short or vanishing delays ($\tau_{\rm delay}\leq0.2$ Gyr) also keep mergers the dominant SMBH growth channel at $z\gtrsim4$ \citep{Fang2025}. 
AzTEC-1's inferred short delay time favors the regime in which $z\approx3$ high-mass analogs contribute to the predicted LISA merger population, and provides some evidence in favor of higher LISA detection counts per year compared to what is expected for longer delay models. 

\subsection{A Recoiling Supermassive Black Hole}
A runaway, or ``recoiling'', SMBH can occur when mass and spin asymmetries produce anisotropic gravitational waves around SMBH binary coalescence, imparting a velocity kick to the remnant \citep{Bekenstein1973}. 
The most convincing recoiling black hole candidates have $\approx1000\,{\rm km\,s^{-1}}$ broad line offsets, comparable to what we measure in AzTEC-1 \citep{Civano2012}. Thus, we now consider the likelihood of the recoiling SMBH scenario. 
\cite{Barrows2025} summarize four observational signatures that a recoiling SMBH scenario could exhibit: (1) a traditional BLR with velocity widths exceeding $>1000\,{\rm km\,s^{-1}}$ \citep[e.g.,][]{Gaskell2009}, (2) a velocity offset between the BLR and systemic velocity of the galaxy \citep[e.g.,][]{Volonteri2008,Bogdanovic2009}, (3) a spatial offset from the host galaxy \citep{Blecha2019} and/or (4) a spatial offset from the NLR \citep{Blecha2013}. 

AzTEC-1 satisfies the kinematic signatures with a broad ($\sim2500\,{\rm km\,s^{-1}}$) H$\alpha$ component offset from the host galaxy's systemic velocity by $-1245\,{\rm km\,s^{-1}}$. The distribution in velocity offsets for recoiling black holes in quasars is estimated to be $<4\%$ for kicks above $500\,{\rm km\,s^{-1}}$ \citep{Bonning2007,Blecha2008}, but this is highly sensitive to the SMBH spin vectors and mass ratio which can maximally produce recoil velocities up to $\sim4000\,{\rm km\,s^{-1}}$ \citep{Campanelli2007}. 
It is also possible that the system is comprised of three massive black holes as found in \cite{Ubler2026} at $z\sim5$, which could readily reproduce escape velocities of $\sim1000\,{\rm km\,s^{-1}}$ \citep{Valtonen1976}. Deeper medium resolution spectra would be needed to test this hypothesis further and search for additional broad components. \cite{Chiaberge2017} find a radio-loud AGN with an offset BLR of $\sim2000\,{\rm km\,s^{-1}}$ that is spatially separated by 11 kpc from the center of the host galaxy. 
The NIRCam imaging data does not reveal any extended structure, but without medium band observations the maps are dominated by stellar continuum. 
The velocity vector of the putative recoiling SMBH could be angled towards the line of sight, such that the observed velocity offset could be close to intrinsic and little angular separation would be observed. Alternatively, AzTEC-1's recoiling SMBH might not have left the central $\sim200$ pc in which case the kick would have had to occurred within $\sim10^5$ yr. This is $\approx100\times$ less than the timescale at which a binary can spend below the final parsec and is accordingly less likely. 

\section{Summary and Conclusions\label{sec:conc}}
We present evidence that broad, blueshifted H$\alpha$ detected at $z_{\rm spec}=4.3419$ in the most luminous millimeter source in the COSMOS field, AzTEC-1 \citep{Smolcic2011,Yun2015}, possibly arises from a sub-pc binary supermassive black hole (SMBH) system, or a recoiling SMBH having received a significant kick $\sim1000\,\rm km\,s^{-1}$, exceeding the escape velocity of the host galaxy and its dark matter halo. An ionized outflow is disfavored by a point-like G395M 2D spectrum, high angular resolution ALMA observations of dust and multiphase line diagnostics with maximum velocities $<300\,{\rm km\,s^{-1}}$ \citep{Iono2016,Tadaki2019,Tadaki2020}, and by energetic arguments. 

From the broad velocity component of H$\alpha$ we measure a black hole mass of $\log\,M_{\rm BH}/M_\odot=8.2\pm0.5$. If AzTEC-1 hosts an equal mass binary or a runaway SMBH, it would fall on the local SMBH mass to stellar mass relation. For a mass ratio of $q=0.1$ AzTEC-1 would be in agreement with $z=6-7$ quasars. 
Most of the stellar mass was formed between $z=4.6-5.8$, $\approx100-400$ Myr prior to $z_{\rm spec}$. If this past burst was driven by a merger that brought in the secondary SMBH, the dynamical crossing time in the inner kpc is $t_{\rm dyn}\approx1$ Myr, sufficiently short to have erased any stellar morphological signatures of the merger event. The presence of significant gas clumps in AzTEC-1's inner few hundred parsec may have formed out of its dynamically unstable disk and played a role extending the delay time between the galaxy merger and SMBH coalescence from $\mathcal{O}(10\,\rm Myr)$ \citep[e.g.,][]{Khan2016} to $\mathcal{O}(100\,\rm Myr)$ \citep[e.g.,][]{Roskar2015}. 
This favors efficient delay times of comparable order in the most massive merging SMBHs at $z>3$. Models with short delay times ($\leq0.2$ Gyr) predict substantially higher SMBH merger rates at $z \gtrsim 3$ than Hubble-time-scale delays, which suppress the high-redshift rate by more than an order of magnitude (and to $\sim0$ by  $z\sim5$, \citealt{Fang2025,Singh2026}). AzTEC-1's inferred $\tau_{\rm delay}\sim 0.1-0.4$ Gyr therefore favors the regime in which such $z\gtrsim3$ high-mass systems contribute detectable LISA events, as well as higher source counts in LISA's first year that longer delay times would suppress. 

Confirming the nature of AzTEC-1's blueshifted broad H$\alpha$ as evidence for a binary/recoiling SMBH host or an outflow requires deeper spatially resolved observations of several optical emission lines, combined with sufficient resolving power ($R\sim1000$) to search for a broad component in [\ion{O}{3}]$\lambda5007$\AA\ and H$\beta$, or a change in the velocity offset of broad H$\alpha$ over the binary orbital cycle. This could be achieved through NIRSpec M grating spectra with either the Integral Field Unit or via slit stepping the MOS. Additionally, AzTEC-1 has a 20 cm flux of $42\pm10\,\mu$Jy \citep{Smolcic2011} which could be resolved at $\approx30$ pc resolution with deep VLBA observations \-- too low to consider resolving anything on the scales of the binary SMBH system; but combined with the ngVLA in the future, can be leveraged for multi-epoch monitoring that has proved successful in selecting robust sub-pc binary candidates at lower redshifts \citep{ONeill2022}. 

Searches for asymmetric broad line profiles using \textit{JWST} NIRSpec have already been used to select candidate SMBH binaries \citep[e.g.,][]{Maiolino2024}. As demonstrated in this work, targeting IR-luminous galaxies with the MSA can be a promising path forward towards efficiently finding more candidates, especially among the most massive binaries at high-redshift that might be detectable in the low-frequency limit of LISA \citep{Katz2019}. The EMBER (\#7076, PI Akins) NIRSpec program has taken G395M spectrum for a handful of other sub-mm bright ($S_{850}\gtrsim1$ mJy, SMGs) galaxies out of $\sim700$ in the COSMOS-Web area \citep{Simpson2019,McKinney2025}. A broader spectroscopic search is needed to uncover more objects like AzTEC-1 and survey the high-redshift landscape of SMBH binaries from an electromagnetic perspective. The redshift distribution of SMGs likely host to massive SMBHs extends to $z\lesssim1$ \citep[e.g.,][]{Brisbin2017,Dudzeviciute2020,Simpson2020} where the stochastic gravitational wave background reported by PTAs is expected to originate \citep{Inayoshi2018,Agazie2023,EPTA2024}.
Such spectroscopic catalogs will therefore be complementary to future space-based gravitational wave catalogs expected from LISA as well as current constraints from PTAs.

\vspace{10pt}
{\small
JM is grateful to Ranga Ram Chary for helpful conversations.  We thank the referee for their helpful and insightful suggestions. 
JM also acknowledges the invaluable labor of the maintenance and clerical staff at our institutions, whose contributions make our scientific discoveries a reality. JM thanks NASA and acknowledges support through the Hubble Fellowship Program, awarded by the Space Telescope Science Institute, which is operated by the Association of Universities for Research in Astronomy, Inc., for NASA, under contract NAS5-26555. HA and ARG acknowledge support by the National Science Foundation Graduate Research Fellowship under grant number DGE 2137420. ORC acknowledges support from National Science Foundation Astronomy \& Astrophysics Postdoctoral Fellowship Award No. 2503202. JBM acknowledges support from NSF Grants AST-2307354 and AST-2408637, and by the NSF-Simons AI Institute for Cosmic Origins. MBK acknowledges support from NSF grant AST-2408247; HST-GO-16686, HST-AR-17028, JWST-GO-03788, and JWST-AR-06278 from the Space Telescope Science Institute, which is operated by AURA, Inc., under NASA contract NAS5-26555; and from the Samuel T. and Fern Yanagisawa Regents Professorship in Astronomy at UT Austin.

Authors from UT Austin acknowledge that UT is an institution that sits on indigenous land. The Tonkawa lived in central Texas, and the Comanche and Apache moved through this area. We pay our respects to all the American Indian and Indigenous Peoples and communities who have been or have become a part of these lands and territories in Texas. We are grateful to be able to live, work, collaborate, and learn on this piece of Turtle Island. 

This work is based [in part] on observations made with the NASA/ESA/CSA James Webb Space Telescope. The data were obtained from the Mikulski Archive for Space Telescopes at the Space Telescope Science Institute, which is operated by the Association of Universities for Research in Astronomy, Inc., under NASA contract NAS 5-03127 for JWST. These observations are associated with programs GTO \#7076, \#7417, \#1727, \#1286, and GO \#5893. 

This paper makes use of the following ALMA data: ADS/JAO.ALMA:2012.1.00978.S, 2015.1.01345.S, 2017.1.00127.S. ALMA is a partnership of ESO (representing its member states), NSF (USA) and NINS (Japan), together with NRC (Canada), MOST and ASIAA (Taiwan), and KASI (Republic of Korea), in cooperation with the Republic of Chile. The Joint ALMA Observatory is operated by ESO, AUI/NRAO and NAOJ. The National Radio Astronomy Observatory is a facility of the National Science Foundation operated under cooperative agreement by Associated Universities, Inc.

Generative AI tools (Claude Opus v$4.8$) were used to spellcheck this manuscript, support literature review, and suggest edits at the final draft stages for streamlining and clarity. 
}

\bibliography{references}{}
\bibliographystyle{aasjournal}

\end{document}